\documentclass[conference]{IEEEtran}
\ifCLASSINFOpdf
\else
\fi
\makeatletter
\@ifundefined{labelindent}{}{\let\labelindent\relax}
\makeatother

\usepackage{color}
\usepackage{amsmath}

\usepackage{comment}
\usepackage{tikz}
\usepackage{amsfonts}
\usepackage{fontawesome}

\usepackage{multirow}
\usepackage{graphicx}
\usepackage{enumitem}
\usepackage{subcaption} 
\usepackage{pifont}
\usepackage{xspace}
\usepackage{tabularx}
\usepackage{caption}
\usepackage{makecell}
\usepackage{booktabs}
\usepackage{array}
\usepackage{colortbl}

\usepackage{mathtools}
\usepackage{backnaur}
\usepackage{amssymb}
\usepackage{amsthm}
\usepackage[breaklinks,hidelinks]{hyperref}
\usepackage{multirow, makecell, multicol}
\usepackage{booktabs}
\usepackage{xspace}
\usepackage{fancyvrb}
\usepackage{siunitx}
\usepackage[normal]{threeparttable}

\usepackage{wasysym}
\usepackage{tikz}
\usetikzlibrary{shapes,arrows}
\usepackage{comment}
\usepackage{scalerel}
\usepackage{xcolor, colortbl}
\usepackage[figuresright]{rotating}

\usepackage{colortbl}
\usepackage{textcomp}  

\DeclareRobustCommand*\circled[1]{\tikz[baseline=(char.base)]{ \node[shape=circle,draw,color=white,fill=black,inner sep=0.5pt] (char){#1};}}

\usepackage[scaled=.7]{beramono}

\newcommand{\system}{\mbox{\textsc{\small{ProTST}}}\xspace}
\newcommand{\systemzero}{\mbox{\textsc{\small{ProTST$_{0}$}}}\xspace}

\def\ie{{i.e.},~}
\def\eg{{e.g.},~}

\usepackage{microtype}

\PassOptionsToPackage{table}{xcolor}
\usepackage{xspace}

\newcommand{\eb}{\begin{equation}}
\newcommand{\ee}{\end{equation}}

\newcommand{\berkay}[1]{{\color{blue}{\bf Berkay:} #1}}
\newcommand{\hanxiao}[1]{{\color{purple}{\bf Hanxiao:} #1}}

\newcommand{\shortsectionBf}[1]{\vspace{-3pt}
\noindent {\bf #1}}
\newcommand{\shortsectionEmph}[1]{\vspace{-1pt}
\noindent {\em #1}}

\begin{document}
%
\title{
\small Extended version of A Progressive Transformer for Unifying Binary Code Embedding and Knowledge Transfer \\(Accepted to IEEE International Conference on Software Analysis, Evolution and Reengineering (SANER), 2025) \\[2ex]
\LARGE \textbf{A Progressive Transformer for Unifying Binary Code Embedding and Knowledge Transfer}
}
\author{
    Hanxiao Lu$^{1}$, Hongyu Cai$^{2}$, Yiming Liang$^{2}$, Antonio Bianchi$^{2}$, Z. Berkay Celik$^{2}$ \\
    \textit{$^1$Columbia University, hl3424@columbia.edu} \\
    \textit{$^2$Purdue University, \{hongyu, liang328, antoniob, zcelik\}@purdue.edu} 
}
\maketitle

\begin{abstract}
    Language model approaches have recently been integrated into binary analysis tasks, such as function similarity detection and function signature recovery. These models typically employ a two-stage training process: pre-training via Masked Language Modeling (MLM) on machine code and fine-tuning for specific tasks. While MLM helps to understand binary code structures, it ignores essential code characteristics, including control and data flow, which negatively affect model generalization. Recent work leverages domain-specific features (e.g., control flow graphs and dynamic execution traces) in transformer-based approaches to improve binary code semantic understanding. However, this approach involves complex feature engineering, a cumbersome and time-consuming process that can introduce predictive uncertainty when dealing with stripped or obfuscated code, leading to a performance drop.
In this paper, we introduce \system, a novel transformer-based methodology for binary code embedding. \system employs a hierarchical training process based on a unique tree-like structure, where knowledge progressively flows from fundamental tasks at the root to more specialized tasks at the leaves. This progressive teacher-student paradigm allows the model to build upon previously learned knowledge, resulting in high-quality embeddings that can be effectively leveraged for diverse downstream binary analysis tasks.
The effectiveness of \system is evaluated in seven binary analysis tasks, and the results show that \system yields an average validation score (F1, MRR, and Recall@1) improvement of $14.8\%$ compared to traditional two-stage training and an average validation score of $10.7\%$ compared to multimodal two-stage frameworks.  

\end{abstract}

\section{Introduction}
\label{sec: intro}
Deep learning, particularly NLP-inspired models, such as RoBERTa~\cite{liu2019roberta} and GPT-based transformers~\cite{jiang2023nova}, have significantly affected diverse downstream binary analysis  tasks, \eg function similarity detection, indirect call recognition, and function signature recovery.
These models leverage transformer architectures, which effectively capture contextual information and long-range dependencies in sequential binaries. The adoption of such models in learning-based binary analysis has opened new avenues to improve the accuracy and efficiency of these tasks~\cite{pei2020xda, jin2022symlm, pei2020trex, pei2021stateformer}.


These models typically follow a two-stage training process: ($1$) an initial pre-training phase via Masked Language Modeling (MLM) on machine code and ($2$) a subsequent fine-tuning phase for a specific task. 
In the pre-training phase, the model learns general representations of the code by predicting masked tokens, which helps to understand the structure and patterns within binary code. 
The fine-tuning phase, on the other hand, involves training the model on a labeled dataset specific to the target task, allowing it to adapt its learned representations to that particular downstream task. 
Fig.~\ref{fig: compare}(a) shows two recent approaches, XDA~\cite{pei2020xda} and Binprov~\cite{he2022binprov}, which use this strategy for disassembly and compiler provenance tasks.
However, in this context, MLM only analyzes the sequential order and relationships between tokens, neglecting binary code's inherent structure and knowledge, such as control and data flow~\cite{wang2022jtrans, zhu2023ktrans}. Recent studies~\cite{pei2020trex, li2021palmtree, wang2022jtrans, jin2022symlm} showed that, without a mechanism to incorporate such structure and knowledge, the accuracy of target tasks may drop.

To address this issue, recent transformer-based binary analysis frameworks~\cite{li2021palmtree, wang2022jtrans, zhu2023ktrans, pei2020trex, pei2021stateformer, pei2022neudep}, as shown in Fig~\ref{fig: compare}(b), incorporate a wide range of domain-specific knowledge into the traditional two-stage architecture by concatenating different high-level modalities as input, \eg assembly language, control flow graphs (CFG), data flow graphs (DFG), and dynamic execution traces. 
%
By integrating these additional features, the model gains a richer understanding of binary code semantics; thus, it may improve the performance of downstream tasks.

Extracting these features, however, necessitates sophisticated reverse engineering tools and specialized knowledge for each CPU architecture. 
This complexity makes them cumbersome to use and limits their effectiveness in diverse binary datasets. 
Furthermore, popular reverse engineering tools struggle with stripped or obfuscated code, and misidentify function boundaries and assembly instructions~\cite{bao2014byteweight}. 
These errors introduce noise and inaccuracies into the extracted features, ultimately degrading the data quality fed to deep learning models.
Moreover, some approaches~\cite{zhu2023ktrans} indiscriminately incorporate a wide array of high-level features, \eg operand type, operand read/write status, and FLAGS register status during pre-training, regardless of the target tasks the model will be fine-tuned for later. 
%
This indiscriminate incorporation can include information not relevant to the specific downstream task. 

These limitations raise a critical research question: \textit{Can we effectively capture the knowledge inherent in binary code without relying on complex reverse engineering and feature engineering, and instead, introduce an appropriate amount of task-specific knowledge, avoiding the disadvantages of including possible modalities into the model during pre-training?}
%
To answer this question, in this paper, we introduce \system, a Progressive Teacher-Student Binary Analysis framework that transfers knowledge between binary analysis tasks to develop high-quality embeddings. We observe that inherent binary knowledge does not necessarily need to be learned from high-level modalities but can be effectively captured through a step-by-step progression of causally related binary tasks. 


To achieve this, \system adopts a hierarchical tree structure, where knowledge progressively flows from fundamental tasks at the root to more specialized tasks at the leaves. That is, lower-level tasks, such as instruction and function boundary recovery, reside near the root, while more specialized tasks, such as function similarity detection and function name prediction, are placed toward the leaves. Each node in this structure functions as a student relative to its predecessor node. The model acquires foundational knowledge from its teacher node and enhances it with task-specific insights before teaching its downstream node, which addresses a more challenging binary task. For example, the function boundary recovery task acts as a student relative to the instruction boundary recovery task. Once it learns from its teacher, it becomes the teacher of the function signature prediction task, which enables the model to leverage the extracted function boundaries for more accurate predictions. The knowledge of the function signature then becomes a teacher for more advanced tasks, such as function similarity detection and function name prediction. 
%

This hierarchical design ensures a natural and logical progression of knowledge, following the teacher-student learning paradigm~\cite{lopez2015unifying, hinton2015distilling}. By placing related tasks with direct logical connections close to each other, the model benefits from a coherent flow of information. 
Unlike traditional two-phase training architectures~\cite{pei2020trex, zhu2023ktrans, pei2020trex, pei2021stateformer, pei2022neudep}, where MLM pre-training is the only strategy irrespective of the downstream task, our progressive teacher-student framework systematically guides the model from basic to advanced tasks. This step-by-step progression, facilitated by the continuous transfer of knowledge through model weight refinement, generates semantically rich embeddings that are fine-tuned for a variety of binary analysis tasks. To our best knowledge, \system is the first work to build knowledge transfer between binary analysis tasks within learning-based binary code embedding.





We evaluate \system on seven binary analysis tasks with diverse datasets. We then compare its performance with state-of-the-art methods in the same experimental setting. Our experiment reveals significant advantages of the teacher-student paradigm compared to a traditional two-stage training architecture. When the teacher-student approach is introduced, we observe an average $14.8\%$ with at least $5\%$ improvement in validation score (F1, MRR, and Recall@1) in all tasks and an average $3$X times faster in convergence. 
Furthermore, we demonstrate improvements in binaries compiled using various optimization settings and dealing with obfuscated data. Compared with multi-modal two-stage frameworks, \system outperforms these models by an average of $10.7\%$ validation score in all tasks.



In summary, we make the following contributions.

\begin{itemize}


    \item We introduce a progressive teacher-student paradigm for binary analysis tasks. This paradigm facilitates efficient knowledge transfer from fundamental tasks to more complex ones, enabling the model to progressively build a hierarchical understanding of binary code.

    \item To realize this paradigm, we introduce a model architecture consisting of three key components: (1) an embedding module that transforms raw binary code into a high-dimensional representation, (2) a transformer backbone model that captures code features, and (3) task-specific heads designed for various objectives. 

    \item We evaluate \system on seven diverse binary analysis tasks. Our results demonstrate that \system yields an average validation score improvement of $14.8\%$ compared to traditional two-stage training and an average validation score of $10.7\%$ compared to multi-modal two-stage frameworks. 

\end{itemize}

\system is publicly available at~\cite{lu2024protst} for use and validation.

\begin{table*}[t]
\centering
\caption{Properties of existing Binary Code Embedding (BCE) approaches and their comparison with \system.}
\label{tab:prevstudy}
\begin{threeparttable}
\resizebox{\textwidth}{!}{%
\begin{tabular}{llcccccccclccccccccc}
\hline
 &  & \multicolumn{8}{c}{Transformer-based} &  & \multicolumn{8}{c}{RNN/CNN/GNN-based} & \multicolumn{1}{l}{} \\ \cline{3-10} \cline{12-19}
 &  & \rotatebox[origin=l]{0}{XDA} & \rotatebox[origin=l]{0}{Binprov} & \rotatebox[origin=l]{0}{Palmtree} & \rotatebox[origin=l]{0}{jTrans} & \rotatebox[origin=l]{0}{kTrans} & \rotatebox[origin=l]{0}{Trex} & \rotatebox[origin=l]{0}{Symlm} & \rotatebox[origin=l]{0}{BinBert} &  & \rotatebox[origin=l]{0}{Malconv2} & \rotatebox[origin=l]{0}{DeepDi} & \rotatebox[origin=l]{0}{Bi-RNN} & \rotatebox[origin=l]{0}{EKLAVYA } & \rotatebox[origin=l]{0}{o-glasses} & \rotatebox[origin=l]{0}{IMCFN} & \rotatebox[origin=l]{0}{SAFE} & \rotatebox[origin=l]{0}{Gemini} & \rotatebox[origin=l]{0}{\system} \\
 &  & \cite{pei2020xda} & \cite{he2022binprov} & \cite{li2021palmtree} & \cite{wang2022jtrans} & \cite{zhu2023ktrans} & \cite{pei2020trex} & \cite{jin2022symlm} & \cite{artuso2024binbert} &  & \cite{raff2021classifying} & \cite{yu_deepdi_2022} & \cite{shin2015recognizing} & \cite{chua2017neural} & \cite{otsubo2020glasses} & \cite{vasan2020imcfn} & \cite{massarelli2019safe} &  \cite{xu2017neural} & $\bigstar$ \\ \hline
 & Raw Bytes & \newmoon & \newmoon & $\cdot$ & $\cdot$ & $\cdot$ & $\cdot$ & $\cdot$ & $\cdot$ &  & \newmoon & $\cdot$ & \newmoon & $\cdot$ & \newmoon & \newmoon & $\cdot$ & $\cdot$ & \newmoon \\
 & Assembly & $\cdot$ & $\cdot$ & \newmoon & \newmoon & \newmoon & \newmoon & \newmoon & \newmoon &  & $\cdot$ & \newmoon & $\cdot$ & \newmoon & $\cdot$ & $\cdot$ & \newmoon & \newmoon & $\cdot$ \\
 & CFG & $\cdot$ & $\cdot$ & \newmoon & \newmoon & $\cdot$ & $\cdot$ & $\cdot$ & $\cdot$ &  & $\cdot$ & $\cdot$ & $\cdot$ & $\cdot$ & $\cdot$ & $\cdot$ & $\cdot$ & \newmoon & $\cdot$ \\
Input Modality & DFG & $\cdot$ & $\cdot$ & \newmoon & $\cdot$ & $\cdot$ & $\cdot$ & $\cdot$ & $\cdot$ &  & $\cdot$ & $\cdot$ & $\cdot$ & $\cdot$ & $\cdot$ & $\cdot$ & $\cdot$ & $\cdot$ & $\cdot$ \\
 & Register Info & $\cdot$ & $\cdot$ & $\cdot$ & $\cdot$ & \newmoon & $\cdot$ & $\cdot$ & $\cdot$ &  & $\cdot$ & $\cdot$ & $\cdot$ & $\cdot$ & $\cdot$ & $\cdot$ & $\cdot$ & $\cdot$ & $\cdot$ \\
 & Operand Info & $\cdot$ & $\cdot$ & $\cdot$ & $\cdot$ & \newmoon & $\cdot$ & $\cdot$ & $\cdot$ &  & $\cdot$ & $\cdot$ & $\cdot$ & $\cdot$ & $\cdot$ & $\cdot$ & $\cdot$ & $\cdot$ & $\cdot$ \\
 & Caller/Callee Info & $\cdot$ & $\cdot$ & $\cdot$ & $\cdot$ & $\cdot$ & $\cdot$ & \newmoon & $\cdot$ &  & $\cdot$ & $\cdot$ & $\cdot$ & $\cdot$ & $\cdot$ & $\cdot$ & $\cdot$ & $\cdot$ & $\cdot$ \\
 & Dynamic Behavior & $\cdot$ & $\cdot$ & $\cdot$ & $\cdot$ & $\cdot$ & \newmoon & \newmoon & \newmoon &  & $\cdot$ & $\cdot$ & $\cdot$ & $\cdot$ & $\cdot$ & $\cdot$ & $\cdot$ & $\cdot$ & $\cdot$ \\ \hline
 & Disassembly & \newmoon & $\cdot$ & $\cdot$ & $\cdot$ & $\cdot$ & $\cdot$ & $\cdot$ & $\cdot$ &  & $\cdot$ & \newmoon & \newmoon & $\cdot$ & $\cdot$ & $\cdot$ & $\cdot$ & $\cdot$ & \newmoon \\
 & Compiler Provenance & $\cdot$ & \newmoon & $\cdot$ & $\cdot$ & $\cdot$ & $\cdot$ & $\cdot$ & \newmoon &  & $\cdot$ & $\cdot$ & $\cdot$ & $\cdot$ & \newmoon & $\cdot$ & $\cdot$ & $\cdot$ & \newmoon \\
Binary Task & Malware Classification & $\cdot$ & $\cdot$ & $\cdot$ & $\cdot$ & $\cdot$ & $\cdot$ & $\cdot$ & $\cdot$ &  & \newmoon & $\cdot$ & $\cdot$ & $\cdot$ & $\cdot$ & \newmoon & $\cdot$ & $\cdot$ & \newmoon \\
 & Function Signature & $\cdot$ & $\cdot$ & \newmoon & $\cdot$ & \newmoon & $\cdot$ & $\cdot$ & $\cdot$ &  & $\cdot$ & $\cdot$ & $\cdot$ & \newmoon & $\cdot$ & $\cdot$ & $\cdot$ & $\cdot$ & \newmoon \\
 & Function Name & $\cdot$ & $\cdot$ & $\cdot$ & $\cdot$ & $\cdot$ & $\cdot$ & \newmoon & $\cdot$ &  & $\cdot$ & $\cdot$ & $\cdot$ & $\cdot$ & $\cdot$ & $\cdot$ & $\cdot$ & $\cdot$ & \newmoon \\
 & Function Similarity & $\cdot$ & $\cdot$ & \newmoon & \newmoon & \newmoon & \newmoon & $\cdot$ & \newmoon &  & $\cdot$ & $\cdot$ & $\cdot$ & $\cdot$ & $\cdot$ & $\cdot$ & \newmoon & \newmoon & \newmoon \\ \hline
\multicolumn{2}{l}{Equipped with Binary code Knowledge} & $\cdot$ & $\cdot$ & \RIGHTcircle & \RIGHTcircle & \RIGHTcircle & \RIGHTcircle & \RIGHTcircle & \RIGHTcircle &  & $\cdot$ & \RIGHTcircle & $\cdot$ & $\cdot$ & $\cdot$ & $\cdot$ & $\cdot$ & \RIGHTcircle & \newmoon \\
\multicolumn{2}{l}{No Feature Engineering} & \newmoon & \newmoon & $\cdot$ & $\cdot$ & $\cdot$ & $\cdot$ & $\cdot$ & $\cdot$ &  & \newmoon & $\cdot$ & \newmoon & \newmoon & \newmoon & \newmoon & \newmoon & $\cdot$ & \newmoon \\ \hline
\end{tabular}%
}
    \begin{tablenotes}
    \item [\newmoon] This mark denotes the feature is fully implemented in the model.
    \item [\RIGHTcircle] This mark denotes that the model has a partial implementation of this feature.
    \end{tablenotes}
\end{threeparttable}

\end{table*}

\section{Background and Related Work}
\label{sec: related}
Binary Code Embedding (BCE) is a technique that maps raw binary code into a lower-dimensional space, allowing these embeddings to be used for various binary analysis tasks. 
We focus on deep learning-based approaches to BCE, specifically those that could operate directly on raw-byte sequences. Tasks such as vulnerability search~\cite{gao2018vulseeker,luo2023vulhawk,yang2023asteria}, memory dependency analysis~\cite{guo2019deepvsa,pei2022neudep}, variable type recovery~\cite{pei2021stateformer}, indirect call recognition~\cite{zhu2023callee}, and binary code comprehension~\cite{ye2023cp, xiong2023hext5} require assembly or higher-level pseudo-code for analysis and are therefore not within the scope of this paper. 
Below, we present BCE approaches by grouping them into learning-based and transformer-based. Table~\ref{tab:prevstudy} compares key characteristics of recent works using these approaches and \system.

\shortsectionBf{Learning-based BCE.}
With the increasing availability of large datasets and advances in deep learning, most BCE approaches initially used learning-based approaches. These methods can be broadly categorized into three groups: ($1$) Sequence-based, ($2$) CNN-based, and ($3$) GNN-based. 
Sequence-based models such as SAFE~\cite{massarelli2019safe}, EKLAVYA~\cite{chua2017neural}, and Bi-RNN~\cite{shin2015recognizing} use RNNs (\eg LSTMs~\cite{schmidhuber1997long}, GRUs~\cite{chung2014empirical}) for tasks such as code similarity detection, function signature prediction, and boundary detection. 
Malconv2~\cite{raff2021classifying} and o-glasses~\cite{otsubo2020glasses} use 1-d CNNs to capture binary code embeddings for tasks such as malware classification and compiler provenance. 
Additionally, IMCFN~\cite{vasan2020imcfn} takes a different approach, using 2-dimensional CNNs to embed binaries as images specifically for malware classification.
Techniques such as Gemini~\cite{zhu2016gemini}, DeepDi~\cite{yu_deepdi_2022}, and Structure2Vec~\cite{massarelli2019investigating} use GNNs to model binary code using graph representations, \eg CFGs, DFGs. 

Current learning-based methods, however, often neglect the nuances of individual instruction formats and complexities. Their focus on static analysis also limits their ability to incorporate broader contextual information essential for comprehensive binary semantic understanding.



\begin{figure}[th!]
\begin{subfigure}{\columnwidth}
  \centering
  \includegraphics[width=\linewidth]{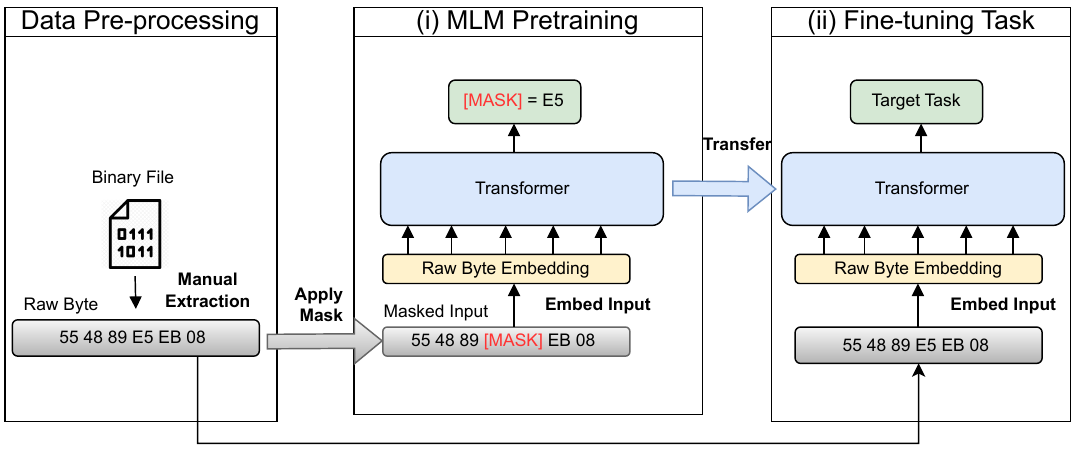}
  \caption{Traditional two-stage architecture.}
  \label{fig:compare_ts}
\end{subfigure}
\begin{subfigure}{\columnwidth}
  \centering
  \includegraphics[width=\linewidth]{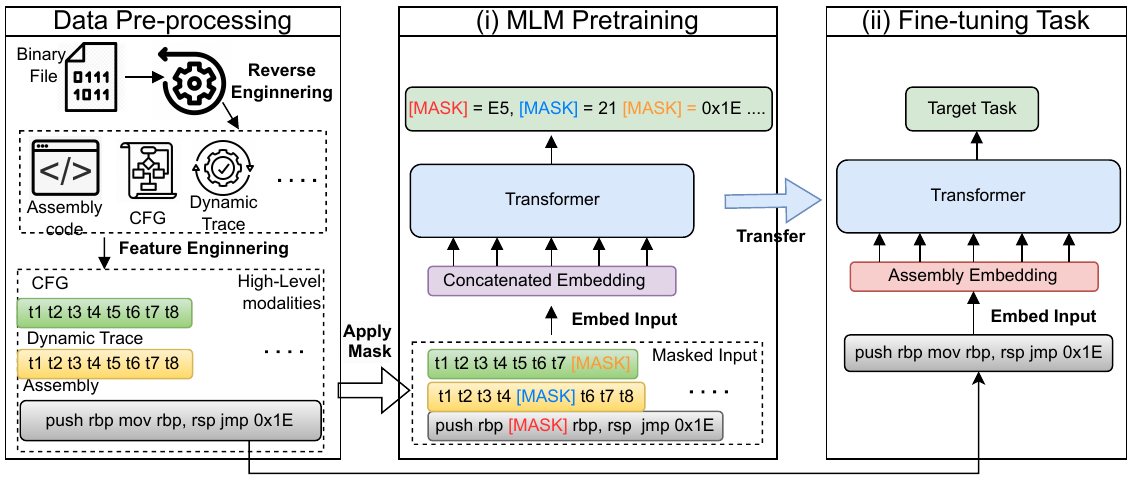}
  \caption{Two-stage framework with multi-modality.}
  \label{fig:compare_mm}
\end{subfigure}
\caption{(a) Traditional two-stage training on static code. (b) Two-stage training with high-level modalities (\eg CFG, DFG, and execution traces). 
}
\label{fig: compare}
\end{figure}

\shortsectionBf{Transformer-based BCE.} Recent advancements in NLP, particularly transformer models, have sparked a series of BCE techniques. 
These methods leverage the transformer's self-attention mechanism to capture long-range dependencies and complex patterns within raw binary code. 
They typically involve a two-stage training process: pre-training with MLM followed by fine-tuning on the specific target task. 
Models such as XDA~\cite{pei2020xda} and BinProv~\cite{he2022binprov} (Figure~\ref{fig: compare}(a)) directly apply transformers to raw byte code for tasks, \eg disassembly and compiler provenance analysis. 
Other models recognize the importance of including the inherent knowledge of binary code to enhance understanding (Figure~\ref{fig: compare}(b)). They achieve this by working on assembly code and embedding information from CFGs and DFGs into the model. This approach is used by Palmtree~\cite{li2021palmtree} and jTrans~\cite{wang2022jtrans} for tasks that include function signature recovery and function similarity detection. 
A line of work, such as Trex~\cite{pei2020trex} and BinBert~\cite{artuso2024binbert}, model the dynamic behavior of binary code within the transformer architecture for tasks, \eg function similarity detection. Other methods, kTrans~\cite{zhu2023ktrans} (using register and operand information) and Symlm~\cite{jin2022symlm} (leveraging caller/callee information), demonstrate the potential of incorporating specific details for tasks such as function name prediction and function similarity detection. 

%
Existing transformer-based methods for binary analysis, however, face two challenges. First, the extraction of features requires specialized knowledge per architecture, limiting their use in diverse datasets. Popular reverse engineering tools face difficulties with obfuscated code and misinterpret functions/instructions, introducing data errors that degrade model performance. Second, indiscriminately incorporating high-level information during pre-training can negatively impact model performance, especially if the information isn't closely relevant to the specific tasks the model will be fine-tuned for later.




\begin{figure*}[th!]
    \centering
    \includegraphics[width=0.82\textwidth]{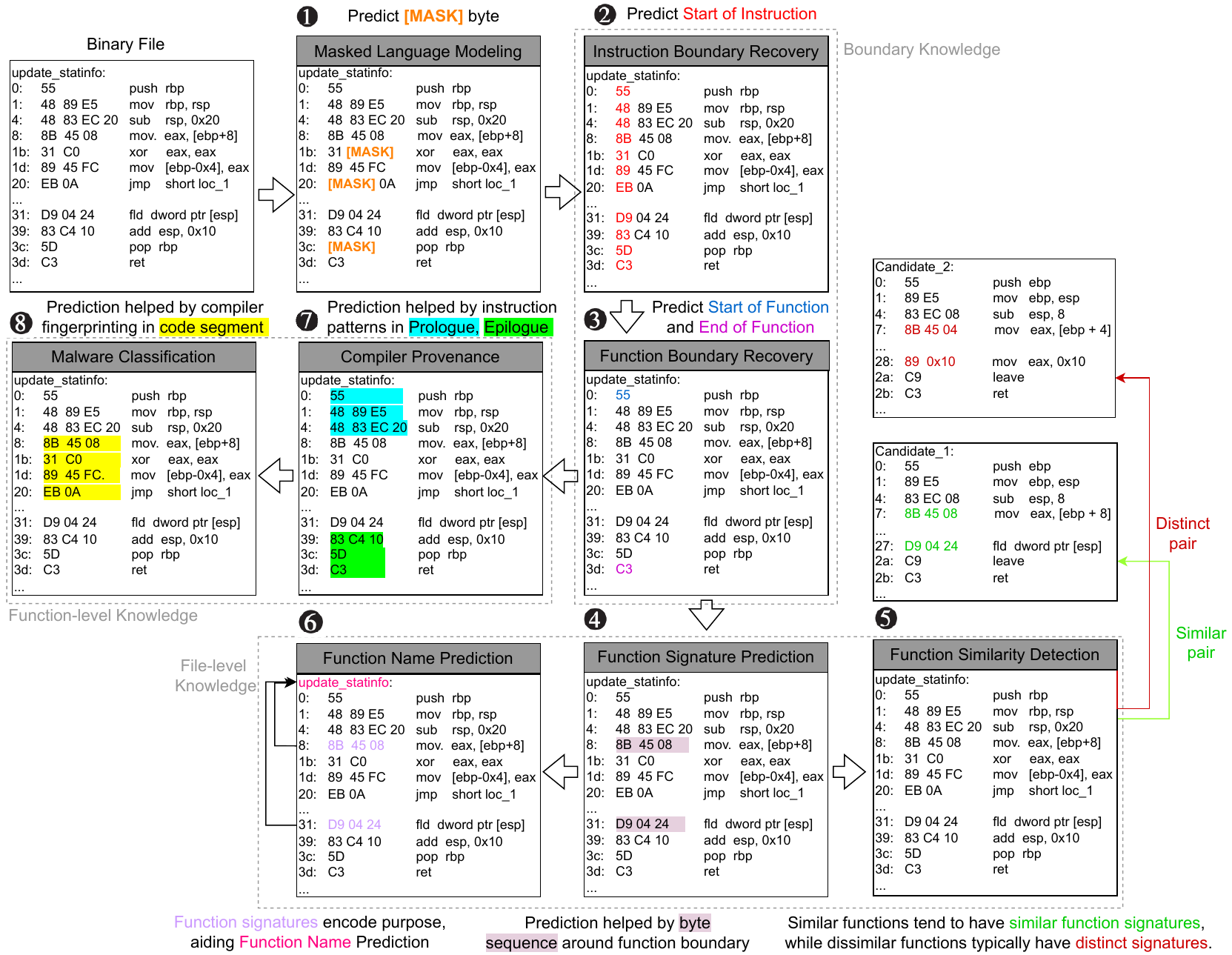}
    \caption{The progressive \emph{teacher-student} learning of \system. Tasks are hierarchically structured to leverage foundational knowledge for more complex tasks. Each task employs a transformer, with model weights serving as interfaces between adjacent nodes. The system operates solely on the raw byte sequence (address and assembly are shown for illustration only). 
    } 
    \label{fig: overview}
\end{figure*}

\section{Teacher-Student Learning for Binary Tasks}
We introduce \system, a novel transformer-based framework for binary analysis tasks, which leverages a teacher-student learning paradigm~\cite{lopez2015unifying,hinton2015distilling}.
\system organizes a series of tasks in progressive order, where each subsequent task builds on the knowledge gained from its predecessors. 

Here, each task specializes in its domain, acting as a focused teacher on the subsequent student task. This progressive knowledge transfer provides a deeper understanding of binary code than cramming various high-level modalities (\eg CFG, execution trace, register information) into a single model. 
In addition, the collaborative nature of \system allows each stage to build upon the knowledge acquired by previous stages. 
This results in an effective transfer of knowledge regarding semantic patterns among tasks, eliminating the need for customized embeddings, complex feature and reverse engineering, or explicit knowledge of the binary code structure.



%

\shortsectionBf{Motivation and Approach Overview.}
Figure~\ref{fig: overview} shows the architecture of \system, which conceptualizes the relationships between binary analysis tasks as a series of teacher-student paradigms. In the following, we will detail how the knowledge acquired and representations learned during the execution of one task (the teacher) can be transferred and leveraged to enhance the performance on a subsequent task (the student). We will empirically validate these relationships in Section~\ref{sec:evaluation}.

\system is composed of two main stages: (a) a Masked Language Modeling (MLM) stage (\circled{\small{1}}) to understand the syntax of the language at the byte level, and (b) a novel Binary Knowledge Accumulation (BKA) stage (\circled{\small{2}}-\circled{\small{8}}) to capture and transfer inherent knowledge within binary code across various tasks.
In the BKA stage, the model learns three types of knowledge: boundary knowledge (\circled{\small{2}}-\circled{\small{3}}), function-level knowledge (\circled{\small{4}}-\circled{\small{6}}), and file-level knowledge (\circled{\small{7}}-\circled{\small{8}}).

The initial MLM stage (\circled{\small{1}}) acts as the first teacher. It trains a base model to predict masked bytes within the binary, equipping the model with a fundamental understanding of the binary's structure and content, akin to learning the alphabet of a new language. 
The student model then addresses the instruction boundary recovery (\circled{\small{2}}), leveraging the foundational knowledge from the MLM stage to identify boundaries between individual instructions. This crucial step enables the model to group bytes into meaningful instruction sequences, similar to parsing words from sentences. 
Building on this foundation, function boundary recovery (\circled{\small{3}}) becomes the next teacher, enhancing the student's ability to delineate functional blocks within the binary based on the previously learned instruction boundaries. Mastering function boundaries is essential for further analysis reliant on the program's functional organization, akin to understanding paragraphs and their purposes within a text.

Once the student model has grasped these two levels of boundary knowledge, it delves deeper into function-level knowledge. This includes function signature prediction (\circled{\small{4}}), which act as high-level summary of a function's purpose, similar to titles for chapters in a book. Accurate predictions offer valuable insights into a function's role within the program. Precise boundary knowledge is crucial for accurate signature prediction. Recognizing where functions start and end helps the model isolate and focus on the relevant bytes that make up a function. This focus ensures that the model analyzes only the pertinent parts of the code, improving its understanding of the function's structure and behavior.

In addition, identifying individual instructions within a function offers a more granular view of its operations. This perspective helps the model determine the number of arguments a function accepts and the data types it returns. In essence, both function and instruction boundary knowledge provide a strong foundation for accurate function signature prediction.  

The extracted semantics from a function's signature form a unique fingerprint. This fingerprint captures a high-level overview of the function's purpose and its interactions within the program. By analyzing and comparing these fingerprints, the model can detect functions with similar functionalities, even if their internal implementations differ. This capability is important for function similarity detection (\circled{\small{5}}). Leveraging these fingerprints facilitates the model's discovery of shared functionalities and code reuse patterns across functions. Furthermore, the patterns derived from function signatures offer valuable clues about a function's purpose. This information is beneficial for the task of function name prediction (\circled{\small{6}}). By recognizing similar signatures, the model can infer the likely role and behavior of a function, guiding it to assign descriptive, human-readable names that reflect each function's purpose.


Beyond understanding the internal structure and functionality of binaries, \system also extends its focus to file-level knowledge, with particular attention to compiler provenance (\circled{\small{7}}) and malware classification (\circled{\small{8}}). Here, the two levels of boundary knowledge acquired during pre-training come into play. The boundary knowledge facilitates the model in detecting distinct patterns and traits associated with specific compilers. Function prologues and epilogues often contain unique sequences of instructions that are characteristic of the compiler used. These, along with other internal instruction patterns, provide valuable clues that help to pinpoint the likely compiler that generated the binary. Understanding compiler-specific traits reveals optimizations and behaviors that influence the binary's performance and structure. In the realm of security, particularly in malware classification, this knowledge is crucial. The design of \system assumes that malware from the same family is often compiled using the same (or a similar) compiler~\cite{gibert2024machine}. Consequently, identifying the compiler used can be instrumental in tracing malware's origins and family.

\begin{figure*}[th!]
    \centering
    \includegraphics[width=0.9\textwidth]{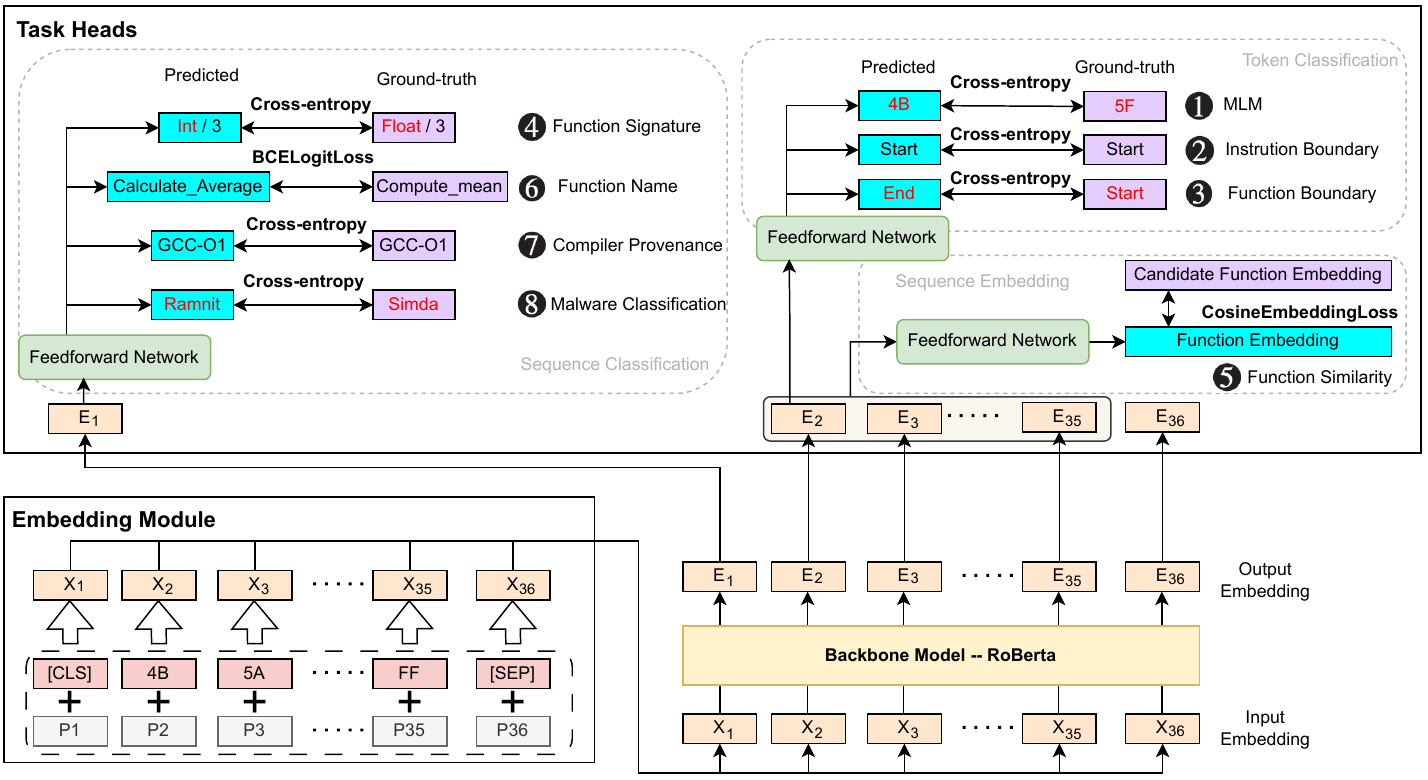}
    \caption{The model architecture of \system} 
    \label{fig: model architecture}
\end{figure*}

\subsection{Model Architecture}
Figure~\ref{fig: model architecture} illustrates the model architecture of \system across all binary analysis tasks. This unified architecture enables the transfer of knowledge and representations between tasks. The model comprises three components: ($1$) embedding module, ($2$) backbone model, and ($3$) task head module. 


\subsubsection{Embedding Module}
The embedding module is the first step for processing binary code. It transforms the raw data of a binary file, a sequence of bytes for further analysis by the model. 
We define the input $x$ as a sequence of byte tokens of size $n$: $x=\{0x00,...,0xff\}^n$. Each input byte $x_i \in x$ is represented as a one-hot encoded vector, \eg $a3$ is encoded as a $256$ dimensional vector with all $0$s but single $1$ at position $163$. 
In addition to the possible $256$ byte values, we add $5$ special tokens to the input vocabulary, including [CLS] at the beginning and [SEP] at the end. [PAD] tokens are appended to the end of the token sequence to ensure equal length for each sequence. Token sequences that exceed the maximum length limit are truncated. For tokens not found in the vocabulary, we uniformly represent them with the special [UNK] token. [Mask] tokens are applied to the input byte token to perform MLM pre-training. 
Notably, we do not impose constraints on the input sequence length $n$. This allows flexibility, enabling byte sequences to span the entire binary program or focus on specific subsets of instructions within a single binary.


Beyond byte-level information, capturing relative positions within the code is crucial to understanding its meaning.
Unlike natural language, where the swapping of two words can roughly preserve the same semantic meaning, swapping two bytes can significantly change the instructions. 
To address this issue, a widely used learned positional encoding method~\cite{liu2019roberta} is employed. This method first transforms the one-hot encoded byte token $x_i$ through an embedding $E_{byte}(x_i)$. This embedding captures the semantic meaning of the individual byte. We then incorporate positional information into the model by applying a learned positional encoding $E_{pos}(i)$ based on the specific position $i$ of the byte token $x_i$. Lastly, a final input embedding is created, $E_i(x_i)$, for the byte token by combining:
\begin{equation}
    E_i(x_i)= E_{byte}(x_i) + E_{pos}(i)
\end{equation}

\subsubsection{Backbone Model}
The backbone model is the core component for processing the embedded binary code generated by the embedding module. We adopt a multi-layer transformer encoder, RoBERTa~\cite{liu2019roberta}, as the backbone model. The transformer is a bidirectional language model based on the self-attention mechanism, which allows it to capture contextual dependencies between tokens (byte embeddings) at different levels of abstraction.

The self-attention mechanism computes a weighted sum of the input embeddings, where the pairwise similarity between the tokens determines the weight. Let \(X = [x_1, x_2, ..., x_n]\) be a sequence of byte embeddings, where \(x_i \in \mathbb{R}^d\) and \(d\) is the embedding dimension. The self-attention mechanism can be expressed as follows:
\begin{equation}
Attention(X) = softmax\left(\frac{XW_q(XW_k)^T}{\sqrt{d_k}}\right)XW_v
\end{equation}

where \(W_q\), \(W_k\), and \(W_v\) are learned weight matrices, and \(d_k\) is a scaling factor. 

In the context of binary analysis, each \(x_i\) represents an embedded byte. The self-attention mechanism allows the model to weigh the importance of each byte in the context of all other bytes in the sequence. This enables the model to learn complex relationships between bytes, even if they are not directly adjacent in the sequence. For instance, the model can learn that a byte representing a function call is related to the bytes representing the corresponding function definition, regardless of their distance in the binary.

Within our framework, we define a sequence of \(T\) binary analysis tasks, with each task \(t \in \{1, 2, ..., T\}\) employing a dedicated instance of the RoBERTa transformer, denoted as \(M_t\). The knowledge transfer process proceeds sequentially through this task chain.
More specifically, we consider \(\Theta_t\) to represent the set of parameters for the model \(M_t\). For the initial task (\(t = 1\)), the model is trained on a labeled dataset \(D_1\), optimizing its parameters to minimize a task-specific loss function \(L_1\):
\begin{equation}
\Theta_1^* = \underset{\Theta_1}{argmin} \ L_1(M_1(\Theta_1), D_1)
\end{equation}

For subsequent tasks (\(t > 1\)), the knowledge acquired by the previous model \(M_{t-1}\) is transferred to initialize the parameters of the current model \(M_t\):
\begin{equation}
\Theta_t^{init} = \Theta_{t-1}^*
\end{equation}

Following this initialization, further fine-tuning is performed on the task-specific dataset \(D_t\), optimizing the parameters to minimize the loss \(L_t\):
\begin{equation}
\Theta_t^* = \underset{\Theta_t}{argmin} \ L_t(M_t(\Theta_t), D_t)
\end{equation}

This iterative process of knowledge transfer and fine-tuning allows each model \(M_t\) to benefit from the knowledge accumulated by its predecessors in the task chain, leading to a progressive refinement of representations and improved performance on downstream tasks.

In contrast to the common practice of freezing certain model parameters during fine-tuning, we allow all parameters in the backbone to be updated during training for each task \(t\)~\cite{min2023recent}. This complete fine-tuning approach provides the model with greater flexibility to adapt to the specific nuances of each task and leads to improved performance (See Section~\ref{sec:evaluation}).

\subsubsection{Task Head Module} 
\label{sec: task head}
The BKA stage of \system encompasses a diverse set of tasks to accumulate knowledge about binary code. 
Each task addresses a specific aspect of binary analysis, and to facilitate this learning process, we equip the backbone model with distinct task-specific ``heads''. The heads serve as interfaces that enable the model to apply the knowledge from the backbone to the specific requirements of each task.


\shortsectionBf{Masked Language Modeling Head.} 
The MLM head (\circled{\small{1}} in Figure~\ref{fig: model architecture}) operates during the initial pre-training stage. It is designed to predict masked bytes within the binary code by leveraging the context from surrounding tokens. This task encourages the model to develop a fundamental understanding of the structure and semantics inherent in the raw byte sequence, akin to learning the vocabulary and grammar of a new language.

We adopt a configurable MLM strategy~\cite{pei2020xda}, where we randomly mask a proportion \(p_{\text{mask}}\) of the input bytes. Of these masked bytes, a fraction \(p_{\text{replace}}\) are replaced with the special [MASK] token, while the remaining \(1 - p_{\text{replace}}\) are replaced with random bytes from the vocabulary \{0x00, ..., 0xff\}.

Formally, let \(x\) denote the original input byte sequence and \(m^x\) the indices of the masked tokens. The masked input sequence \(x^{MLM}\) can be expressed as:
\begin{equation}
        x^{MLM} = REPLACE(x, m^x, [MASK])
\end{equation}

The objective of the MLM head is to reconstruct the original masked bytes, formulated as the following loss function:
\begin{equation}
        \mathcal{L}_{MLM} = - \sum_{i \in m^x} \log P(x_i \mid x^{MLM})
\end{equation}

This objective function drives the model to maximize the probability of reconstructing the original masked bytes to learn meaningful data representations.

\shortsectionBf{Binary Task Heads.} 
Following the MLM pre-training, \system addresses seven distinct binary analysis tasks within the BKA stage. We employ three types of task-specific heads, each tailored to the particular classification problem.

\shortsectionEmph{Sequence-Level Classifier.} We use this head for tasks that necessitate the characterization of an entire byte sequence, such as predicting function names, identifying compilers, or classifying malware families (tasks \circled{\small{4}}, \circled{\small{6}}-\circled{\small{8}}). It operates on the final hidden state \(h_{[CLS]}\) corresponding to the [CLS] token produced by the backbone model.

\shortsectionEmph{Token-Level Classifier.} With this head, we focus on fine-grained analysis of the byte sequence, where the goal is to assign a class label to each individual byte (token). Tasks including instruction and function boundary recovery (tasks \circled{\small{2}}-\circled{\small{3}}) fall into this category. For each token \(x_i\) in the input sequence \(x\), the token-level classifier processes its corresponding hidden state \(h_i\) from the backbone model and produces an output \(y_i\).

\shortsectionEmph{Sequence Embedding Head.} We use this head for the task of function similarity detection (\circled{\small{5}}). It generates a fixed-length embedding representation \(e\) for an entire byte sequence \(x\) by aggregating (e.g., averaging) the output embeddings \(h_1, h_2, ..., h_n\) from the backbone model. These sequence embeddings can then be compared using cosine similarity or other suitable distance metrics to assess the functional similarity between different code segments.

The choice of loss function depends on the specific task. For multi-class, multi-label classification tasks such as function name prediction, we employ the binary cross-entropy loss with logits (BCELogit Loss). For function similarity detection, which involves embedding comparison, we use the Cosine Embedding Loss. For all other standard multi-class classification tasks, the Cross-Entropy Loss is employed. Detailed information on the ground truth for each task and the specific configuration of each task head can be found in Appendix~\ref{sec: head details}.

Although \system involves a multi-stage pre-training process, it utilizes only one instance of RoBERTa during inference. This design allows for efficient adaptation to new tasks. If a new binary analysis task needs to be added to the hierarchy, only the new task after the insertion point requires further training. For instance, if a new task needs to be placed after function boundary recovery, we can leverage the pre-trained checkpoints stored at that stage and continue fine-tuning the new task.

\section{Implementation and Evaluation Setup}
\label{sec:eval-setup}
We evaluate \system using seven binary tasks, each performed on a distinct dataset. Since binaries within a single dataset may exhibit similar patterns, this diversity in data patterns is crucial to demonstrate \system's generality. In this paper, we focus on the x86 architecture, as it is prevalent in BCE research involving multiple tasks~\cite{wang2022jtrans, zhu2023ktrans, li2021palmtree}. The summary of each dataset is provided in Table~\ref{tab: datasets} and their details are given in Appendix~\ref{sec: dataset}.


\subsection{Model Configuration}
\label{sec: training config}

We configure \system with the pre-trained RoBERTa model, using its default settings of $12$ layers, $12$ attention heads per layer, and a hidden dimension of $768$. The maximum input sequence length is set to $512$ tokens.

\shortsectionBf{Pre-training.} In this stage, we employ either a binary or project-level split depending on the task. For instruction/function boundary recovery and malware classification, a binary-level split is used to ensure that the model is evaluated on entirely unseen binaries. For other tasks, a project-level split is used to ensure that data from different projects are kept separate during pre-training and evaluation. 

In both cases, the split ratio is $90$\% for pre-training and $10$\% for evaluation. This setup aims to assess the model's ability to generalize to unseen data and avoid overfitting. During pre-training, each task (teacher) is trained with a batch size of $96$ samples for $20$ epochs. The MLM pre-training task adopts \(p_{\text{replace}}\) of 0.5, which aligns with XDA~\cite{pei2020xda}.

\shortsectionBf{Fine-tuning.} For fine-tuning, we randomly select $100$K samples from the evaluation dataset of each task, except for malware classification, where we use $10$K samples due to limited data availability in BIG2015. The selected samples are then split into fine-tuning and testing sets. For tasks involving instruction boundaries, compiler provenance, function signatures, and function similarity detection, we allocate $1$\% of the samples for fine-tuning and $99$\% for testing. With this train-test ratio, we aim to minimize overfitting, better generalize to unseen data, and expose the pre-trained teacher models to a much larger and more diverse dataset than the student models to enable effective knowledge transfer. A less strict split of $10$\% for fine-tuning and $90$\% for testing is employed for other tasks.  Similar to pre-training, fine-tuning involves processing the data in batches of $96$ samples for $100$ epochs.

\begin{table}[t]
\centering
\caption{Binary datasets used to evaluate \system. 
} 
\setlength{\tabcolsep}{2pt}
\resizebox{\columnwidth}{!}{%
\begin{tabular}{lll}
\hline
\textbf{Dataset} & \textbf{Data Size} & \textbf{Task} \\ \hline
Binutils~\cite{binutils} & 1 project & Masked Language Modeling \\
SPEC CPU~\cite{SPEC_CPU_2017, SPEC_CPU_2006} & 2.01G bytes & Inst./Func. Boundary Prediction \\
BAP~\cite{bao2014byteweight} & 345M bytes & Func. Boundary Prediction \\
BinKit~\cite{kim2022revisiting} & 75M functions & Compiler Prov./Func. Signature Prediction \\
Symlm~\cite{jin2022symlm} & 1.44M functions & Func. Name Prediction\\
Binarycorp-3M~\cite{wang2022jtrans} & 404K functions & Func. Similarity Detection \\
BIG2015~\cite{ronen2018microsoft}& 10860 files & Malware Classification \\ \hline
\end{tabular}%
}
\label{tab: datasets}
\end{table}




\subsection{Evaluation Metrics}
\label{sec: eval metric}
To assess \system's performance in binary analysis tasks, we employ established evaluation metrics from previous work.




Following recent work~\cite{pei2020xda, jin2022symlm, pei2021stateformer}, the primary metric used to evaluate most tasks in \system is the F1 score. To account for potential class imbalances that might skew performance assessments, we specifically employ the macro-F1 score variant. This metric is computed by first calculating the F1-score for each individual class, which is the harmonic mean of precision and recall. Then we average the F1-scores across all classes, which yields the macro-F1 score:
\begin{equation}
\text{Macro-F1} = \frac{1}{|C|} \sum_{c \in C} F1_c
\end{equation}

where \(|C|\) denotes the total number of classes, and \(F1_c\) represents the F1-score for class \(c\). This ensures that all classes contribute equally to the overall evaluation regardless of their frequency in the dataset.

For the task of function similarity detection, we adopt the Mean Reciprocal Rank (MRR) and Recall@k metrics from a recent work~\cite{wang2022jtrans}. 
MRR quantifies how well the model ranks the most similar function (ground truth) relative to others for a given query function:
\begin{equation}
MRR = \frac{1}{|Q|} \sum_{q_i \in Q} \frac{1}{\text{rank}(q_i^{gt})} 
\end{equation}

where \(|Q|\) is the total number of query functions in the evaluation set, \(q_i\) represents a query function, \(q_i^{gt}\) denotes its ground truth counterpart, and \(\text{rank}(q_i^{gt})\) indicates the position of the ground truth function in the ranked list returned by the model for query \(q_i\). A lower rank signifies a better result.

Recall@k, as a complementary metric, measures the proportion of queries in which the ground truth function is included within the top \(k\) results retrieved by the model:
\begin{equation}
Recall@k = \frac{1}{|Q|} \sum_{q_i \in Q} \mathbb{I}(\text{rank}(q_i^{gt}) \leq k)
\end{equation}

where \(\mathbb{I}(\cdot)\) is an indicator function that equals $1$ if the condition inside the parentheses is true and $0$ otherwise.

We note that some tasks within our framework have distinct subtasks. Function signature prediction, for example, includes predicting both argument count and return type. Similarly, function similarity detection involves evaluating the similarity at varying sizes (32 and 10K functions). To fully assess the \system's performance, we report metrics for each subtask.

\section{Evaluation}
\label{sec:evaluation}
We evaluate \system in seven diverse binary analysis tasks using the datasets and evaluation metrics described in Section~\ref{sec:eval-setup}. We present our findings through several key research questions:

\begin{enumerate} [label=\textbf{RQ\arabic*}, leftmargin=9mm, topsep=3pt, itemsep=0.25ex]

\item Does the progressive teacher-student paradigm employed by \system yield improved performance on downstream binary analysis tasks compared to established baselines?

\item To what extent does \system generalize to varying binary optimization levels and code obfuscation techniques?

\item How does the order of tasks in the BKA stage influence knowledge transfer and impact overall performance?

\item What is the computational efficiency of \system compared to alternative approaches, and how does it scale with task complexity and dataset size?
\end{enumerate}

To ensure a fair comparison, in the experiments, all models undergo the same fine-tuning procedure as for \system, detailed in Section~\ref{sec: training config}. All experiments were performed on a dedicated server with eight AMD EPYC 7543 32-core processors, one A100 GPU, 32GB memory, and 1TB SSD. 

\begin{table}[t]
\centering
\caption{Results on instruction boundary recovery.}
\setlength{\tabcolsep}{10pt}
\resizebox{0.75\columnwidth}{!}{%
\begin{tabular}{l|rr|r}
\hline
\multirow{2}{*}{\textbf{Model}} & \multicolumn{2}{c|}{\textbf{Class}} & \multicolumn{1}{l}{\multirow{2}{*}{\textbf{Average}}} \\
 & \multicolumn{1}{l}{\textbf{Start}} & \multicolumn{1}{l|}{\textbf{Middle}} & \multicolumn{1}{l}{} \\ \hline
Bi-RNN~\cite{shin2015recognizing} & 0.443 & 0.885 & 0.664 \\
DeepDi~\cite{yu_deepdi_2022} & 0.765 & \textbf{0.998} & 0.881 \\
\systemzero & 0.61 & 0.906 & 0.758 \\
\system & \textbf{0.827} & 0.941 & \textbf{0.884} \\ \hline
\end{tabular}%
}
\label{tab: RQ1_instbound}
\end{table}

\subsection{RQ1: Overall Binary Task Effectiveness}
\label{sec:RQ1}

We present the performance of \system in each individual task and compare it with the state-of-the-art method of each task. For each task, we report the performance metrics of models for each class within the dataset.

To further investigate the efficacy of the teacher-student paradigm, we introduce a variant model, \systemzero. In contrast to \system, which leverages fine-tuning on student tasks to refine the transferred knowledge, \systemzero directly leverages the embeddings generated by the pre-trained teacher model. This approach aligns with the principles of zero-shot learning~\cite{xian2017zero}, in which a model is evaluated on unseen tasks without any task-specific adaptation. The performance of \systemzero is a critical indicator of knowledge transferability by the teacher model, as it relies solely on the teacher's knowledge to generalize to novel tasks.

\subsubsection{Instruction Boundary Recovery} 
The instruction boundary recovery task aims to classify each byte within a binary file as either ``Start'' (the first byte of an assembly instruction) or ``Middle'' (all other bytes within an instruction). We evaluate the performance of \system against two baselines: the Bidirectional Recurrent Neural Network (Bi-RNN) based method by Shin et al.~\cite{shin2015recognizing} and DeepDi~\cite{yu_deepdi_2022}, a graph neural network (GNN)-based disassembler. It is important to note that DeepDi is a commercial tool, and we leverage its functionality through its provided API without fine-tuning the model under the same experimental setup used for the other methods.

Table~\ref{tab: RQ1_instbound} summarizes the results for this task. Our analysis yields several key observations. First, \system achieves the highest average F1-score of 88.4\%, outperforming both Bi-RNN (66.4\%) and DeepDi (88.1\%). Although DeepDi has been trained on a larger dataset of binaries compared to the fine-tuning setting used for \system, it still exhibits lower performance in identifying Start classes, which represent a minor class within the dataset (DeepDi: 76.5\% vs. \system: 82.7\%). This improvement is due to the knowledge transferred from the MLM pre-training stage to instruction boundary recovery. Second, \systemzero exhibits competitive performance over the Bi-RNN baseline; it yields an average F1-score of 75.8\%. This shows that the RoBERTa-based architecture captures local patterns within binary code, even without task-specific fine-tuning. Lastly, a consistent trend across all models is the performance in classifying Middle bytes compared to Start bytes. This disparity is due to the inherent class imbalance, where middle bytes are significantly more prevalent than start bytes, which poses a challenge for models to learn discriminative features for the less frequent class.

\begin{table}[t!]
\centering
\caption{Results on function boundary recovery.}
\setlength{\tabcolsep}{8pt}
\resizebox{0.8\columnwidth}{!}{%
\begin{tabular}{l|rrr|r}
\hline
\multirow{2}{*}{\textbf{Model}} & \multicolumn{3}{c|}{\textbf{Class}} & \multicolumn{1}{l}{\multirow{2}{*}{\textbf{Average}}} \\
 & \multicolumn{1}{l}{\textbf{Middle}} & \multicolumn{1}{l}{\textbf{Start}} & \multicolumn{1}{l|}{\textbf{End}} & \multicolumn{1}{l}{} \\ \hline
Bi-RNN~\cite{shin2015recognizing} & 0.997 & 0 & 0 & 0.332 \\
DeepDi~\cite{yu_deepdi_2022} & 0.999 & 0.741 & 0.741 & 0.827 \\
\systemzero & 0.997 & 0.066 & 0.04 & 0.368 \\
\system & \textbf{0.999} & \textbf{0.849} & \textbf{0.897} & \textbf{0.915} \\ \hline
\end{tabular}%
}
\label{tab: RQ1_funcbound}
\end{table}

\subsubsection{Function Boundary Recovery}
The function boundary recovery task aims to classify each byte within a binary file as either ``Start of Function'', ``Middle of Function'', or ``End of Function''. We evaluate \system against the same methods used in instruction boundary recovery: Bi-RNN~\cite{shin2015recognizing} and DeepDi~\cite{yu_deepdi_2022}. It is important to acknowledge a limitation in DeepDi's API for this task. It cannot differentiate between Start and End classes. Consequently, we will report the average F1-score for the combined Start/End class for DeepDi.

\begin{table*}[ht]
\centering
\caption{Results on compiler provenance.}
\resizebox{\textwidth}{!}{%
\begin{tabular}{l|rrrrrr|rrrrrr|r}
\hline
\multirow{2}{*}{\textbf{Model}} & \multicolumn{6}{c|}{\textbf{GCC}} & \multicolumn{6}{c|}{\textbf{Clang}} & \multicolumn{1}{l}{\multirow{2}{*}{\textbf{Average}}} \\
 & \multicolumn{1}{l}{\textbf{O0}} & \multicolumn{1}{l}{\textbf{O1}} & \multicolumn{1}{l}{\textbf{O2}} & \multicolumn{1}{l}{\textbf{O3}} & \multicolumn{1}{l}{\textbf{Os}} & \multicolumn{1}{l|}{\textbf{Ofast}} & \multicolumn{1}{l}{\textbf{O0}} & \multicolumn{1}{l}{\textbf{O1}} & \multicolumn{1}{l}{\textbf{O2}} & \multicolumn{1}{l}{\textbf{O3}} & \multicolumn{1}{l}{\textbf{Os}} & \multicolumn{1}{l|}{\textbf{Ofast}} & \multicolumn{1}{l}{} \\ \hline
o-glasses~\cite{otsubo2020glasses} & 0.157 & 0.346 & 0.04 & 0.005 & 0.013 & 0.079 & 0.397 & 0.137 & 0.001 & 0 & 0.221 & 0.157 & 0.129 \\
Binprov~\cite{he2022binprov} & 0.888 & 0.663 & 0.27 & 0.221 & 0.517 & 0.228 & 0.918 & 0.321 & \textbf{0.228} & 0.125 & 0.389 & 0.231 & 0.416 \\
\systemzero & 0.485 & 0.317 & 0.048 & 0 & 0 & 0 & 0.578 & 0.243 & 0.043 & 0.1 & 0 & 0.016 & 0.153 \\
\system & \textbf{0.902} & \textbf{0.713} & \textbf{0.293} & \textbf{0.327} & \textbf{0.587} & \textbf{0.267} & \textbf{0.938} & \textbf{0.425} & 0.183 & \textbf{0.247} & \textbf{0.389} & \textbf{0.249} & \textbf{0.46} \\ \hline
\end{tabular}%
}
\label{tab: RQ1_cmpprov}
\end{table*}

\begin{table*}[th!]
\centering
\caption{Results on malware classification task.}
\resizebox{\textwidth}{!}{%
\begin{tabular}{l|rrrrrrrrr|r}
\hline
 & \multicolumn{9}{c|}{\textbf{Malware Class}} & \multicolumn{1}{l}{} \\
\multirow{-2}{*}{\textbf{Model}} & \multicolumn{1}{l}{{ \textbf{Ramnit}}} & \multicolumn{1}{l}{{ \textbf{Lollipop}}} & \multicolumn{1}{l}{{ \textbf{Kelihos\_ver3}}} & \multicolumn{1}{l}{{ \textbf{Vundo}}} & \multicolumn{1}{l}{{ \textbf{Simda}}} & \multicolumn{1}{l}{{ \textbf{Tracur}}} & \multicolumn{1}{l}{{ \textbf{Kelihos\_ver1}}} & \multicolumn{1}{l}{{ \textbf{Obfuscator.ACY}}} & \multicolumn{1}{l|}{{ \textbf{Gatak}}} & \multicolumn{1}{l}{\multirow{-2}{*}{\textbf{Average}}} \\ \hline
IMCFN~\cite{vasan2020imcfn} & 0.529 & 0.8 & 0.986 & 0.474 & 0.026 & 0.493 & 0.855 & 0.711 & 0.474 & 0.594 \\
Malconv2~\cite{raff2021classifying} & 0.751 & 0.941 & 0.985 & 0.334 & 0.026 & 0.587 & 0.74 & 0.748 & 0.701 & 0.646 \\
\systemzero & 0.596 & 0.73 & 0.855 & 0.251 & 0 & 0.223 & 0.011 & 0.615 & 0.434 & 0.413 \\
\system & \textbf{0.824} & \textbf{0.948} & \textbf{0.986} & \textbf{0.837} & \textbf{0.182} & \textbf{0.795} & \textbf{0.871} & \textbf{0.862} & \textbf{0.905} & \textbf{0.801} \\ \hline
\end{tabular}%
}
\label{tab: RQ1_mlwcls}
\end{table*}

Similar to the instruction boundary recovery task, as shown in Table~\ref{tab: RQ1_funcbound}, all models exhibit near-perfect performance on the dominant Middle class but encounter difficulties in classifying the less frequent Start and End classes due to the inherent class imbalance in the dataset. Yet, \system significantly outperforms both DeepDi and Bi-RNN, achieving an average 91.5\% F1 score. Although DeepDi has been trained on a larger data set of binaries compared to the fine-tuning setting used for \system, it still exhibits lower performance in identifying the Start and End classes (DeepDi: 74.1\% vs. \system: 84.9\% at Start and 89.7\% at End). This improvement is attributed to the additional instruction boundary knowledge transferred from the previous stage, which provides valuable contextual information for function boundary recovery. Moreover, \systemzero, shows a rudimentary ability to classify function boundaries and achieves F1-scores of 6.6\% and 4.0\% for Start and End classes. This surpasses the Bi-RNN baseline, which fails to identify any patterns in these classes; this shows the effectiveness of the RoBERTa-based architecture in capturing function boundaries even without task-specific fine-tuning.

\subsubsection{Compiler Provenance}
The compiler provenance task involves identifying both the compiler (GCC or Clang) and its specific optimization level (O0, O1, $\ldots$ ,Ofast) used to generate a given binary file (see Appendix~\ref{sec: head details} for class descriptions). We evaluate the performance of \system against two established methods: O-glasses~\cite{otsubo2020glasses}, which employs 1D convolutions to analyze raw binary sequences for provenance, and Binprov~\cite{he2022binprov}, which uses a transformer-based architecture with a traditional two-stage training approach that operates directly on raw bytes.

Table~\ref{tab: RQ1_cmpprov} presents the results for this task. \system achieves an average F1-score of 46\%, outperforming both Binprov (41.6\%) and O-glasses (12.9\%). Performance improvement is evident in all classes. To better understand the impact of knowledge transfer within \system, we compare its results with \systemzero, which, despite outperforming O-glasses with an average F1-score of 15.3\%, exhibits a clear weakness in identifying certain compiler-optimization combinations--particularly GCC-O3, GCC-Os, GCC-Ofast, and Clang-Os--where it achieves a zero F1-score. This suggests that fine-tuning is crucial for achieving better generalization. Our results also demonstrate that classifying binaries compiled with Clang is more challenging than those compiled with GCC, as all models exhibit lower performance on Clang-compiled binaries.

\subsubsection{Malware Classification}
In this task, we evaluate \system in a malware classification task to identify the specific malware class within a given binary file. We compare \system against two established baselines: IMCFN~\cite{vasan2020imcfn}, which uses a 2D-CNN to extract image-like features from byte sequences, and Malconv2~\cite{raff2021classifying}, which employs a 1D-CNN to directly extract features from the raw 1D representation of the malware bytes.

As shown in Table~\ref{tab: RQ1_mlwcls}, \system outperforms both IMCFN (59.4\%) and Malconv2 (64.6\%), achieving an average F1-score of 80.1\%. We observe that performance improvement is due to knowledge transfer enabled by the teacher-student paradigm and the powerful representation learning capabilities of the RoBERTa backbone, which together lead to learning discriminative representations for each class. However, the relatively small size (10K samples) of the BIG2015 dataset limits the ability of \systemzero to fully leverage the knowledge transferred from the teacher model without fine-tuning, leading to lower performance than the other models. Notably, all models struggle with the Simda class, as Simda samples comprise only 0.4\% of the total dataset.

\begin{table*}[t]
    \centering
    \setlength{\tabcolsep}{2.5pt}
    \caption{Results on function signature prediction: (Left) argument count prediction and (Right) return type prediction }
    \label{tab:funcsig}
    \begin{minipage}[t]{0.48\textwidth} 
        \centering
        \resizebox{1.0\textwidth}{!}{
        \begin{tabular}{l|rrrrrrr|r}
        \hline
        \multirow{2}{*}{\textbf{Model}} & \multicolumn{7}{c|}{\textbf{Number of Arguments Class}} & \multirow{2}{*}{\textbf{Average}} \\
         & \textbf{0} & \textbf{1} & \textbf{2} & \textbf{3} & \textbf{4} & \textbf{5} & \textbf{others} &  \\ \hline
        EKLAVYA~\cite{chua2017neural} & 0.149 & 0.321 & 0.325 & 0.282 & 0.585 & 0.121 & 0.659 & 0.349 \\
        Palmtree~\cite{li2021palmtree} & 0.431 & 0.475 & 0.393 & 0.266 & 0.117 & 0.141 & 0.557 & 0.34 \\
        \systemzero & 0.321 & 0.541 & 0.421 & 0.313 & 0.601 & 0.055 & 0.826 & 0.439 \\
        \system & \textbf{0.853} & \textbf{0.84} & \textbf{0.755} & \textbf{0.737} & \textbf{0.783} & \textbf{0.436} & \textbf{0.94} & \textbf{0.763} \\ \hline
        \end{tabular}%
        }
    \end{minipage}\hfill
    \begin{minipage}[t]{0.48\textwidth}
        \centering
        \resizebox{1.0\textwidth}{!}{
        \begin{tabular}{l|rrrrrr|r}
        \hline
         & \multicolumn{6}{c|}{\textbf{Return Type Class}} &  \\
        \multirow{-2}{*}{\textbf{Model}} & \textbf{int} & \textbf{char} & \textbf{void} & \textbf{double} & \textbf{bool} & \textbf{others} & \multirow{-2}{*}{\textbf{Average}} \\ \hline
        EKLAVYA~\cite{chua2017neural} & 0.074 & 0.035 & 0.779 & 0 & 0.065 & 0.407 & 0.226 \\
        Palmtree~\cite{li2021palmtree} & 0.291 & 0.102 & 0.758 & 0.511 & 0.202 & 0.248 & 0.352 \\
        \systemzero & 0.015 & 0 & 0.78 & 0.075 & 0 & 0.465 & 0.223 \\
        \system & \textbf{0.549} & \textbf{0.495} & \textbf{0.876} & \textbf{0.694} & \textbf{0.423} & \textbf{0.724} & \textbf{0.626} \\ \hline
        \end{tabular}%
        }
    \end{minipage}
\end{table*}

\begin{table}[t]
\centering
\setlength{\tabcolsep}{14pt}
\caption{Results on function name prediction.}
\resizebox{0.6\columnwidth}{!}{%
\begin{tabular}{l|c}
\hline
\textbf{Model} & \textbf{Average} \\ \hline
Asm2vec~\cite{ding2019asm2vec} & 0.604 \\
Symlm~\cite{jin2022symlm} & 0.778 \\
\systemzero & 0.698 \\
\system & \textbf{0.835} \\ \hline
\end{tabular}%
}
\label{tab: RQ1_funcname}
\end{table}

\begin{table*}[t]
\centering
\caption{Results on function similarity detection for pool sizes 32 and 10k.}
\setlength{\tabcolsep}{12pt}
\label{tab:funcsim}
\resizebox{\textwidth}{!}{%
\begin{tabular}{l|rrrrrr|r}
\hline
 & \multicolumn{6}{c|}{\textbf{MRR}} &  \\
\multirow{-2}{*}{\textbf{Model}} & \textbf{O0,O3} & \textbf{O1,O3} & \textbf{O2,O3} & \textbf{O0,Os} & \textbf{O1,Os} & \textbf{O2,Os} & \multirow{-2}{*}{\textbf{Average}} \\ \hline
Asm2vec~\cite{ding2019asm2vec} & 0.212/0.011 & 0.429/0.122 & 0.599/0.296 & 0.215/0.009 & 0.414/0.128 & 0.461/0.157 & 0.422/0.121 \\
Jtrans~\cite{wang2022jtrans} & 0.762/\textbf{0.267} & 0.911/0.611 & \textbf{0.976}/0.762 & 0.815/\textbf{0.323} & 0.926/\textbf{0.631} & 0.934/\textbf{0.625} & 0.887/\textbf{0.536} \\
\systemzero & 0.537/0.13 & 0.857/0.359 & 0.948/0.672 & 0.574/0.146 & 0.823/0.326 & 0.852/0.389 & 0.766/0.337 \\
\system & \textbf{0.796}/0.218 & \textbf{0.943}/\textbf{0.619} & 0.973/\textbf{0.792} & \textbf{0.84}/0.256 & \textbf{0.944}/0.592 & \textbf{0.941}/0.614 & \textbf{0.906}/0.515 \\ \hline
\end{tabular}%
}
\end{table*}

\subsubsection{Function Signature Prediction}
The function signature prediction task involves predicting two crucial aspects of a function's signature: ($a$) the return type and ($b$) the number of arguments it accepts. We evaluate \system against two established methods: EKLAVYA~\cite{chua2017neural}, which uses RNNs to learn function type signatures, and Palmtree~\cite{li2021palmtree}, which leverages a transformer-based architecture that incorporates data/control flow information on assembly code.

Table~\ref{tab:funcsig} summarizes the results. Our analysis reveals that \system outperforms both EKLAVYA and Palmtree in both number of arguments prediction, 76.3\% vs. 34.9\% and 34.0\%, and return type prediction, 62.6\% vs. 22.6\% and 35.2\%. While Palmtree reports high accuracy in its experimental setting (237 binaries for training and 14 binaries for testing), it demonstrates less adaptability to our more challenging fine-tuning setting. This suggests that the domain-specific knowledge incorporated by PalmTree through context window prediction (CWP) and def-use prediction (DUP) tasks is less effective than the knowledge transfer achieved through \system's teacher-student paradigm.
Furthermore,  \systemzero, performs competitively, matching EKLAVYA's performance in return-type prediction (around 22\% for both) and surpassing EKLAVYA and Palmtree in number of arguments prediction (43.9\% vs. 34.9\% and 34.0\%). However, \systemzero fails to distinguish certain less frequent return-type classes than EKLAVYA, particularly for bool and char.

\subsubsection{Function Name Prediction}
The function name prediction task involves assigning human-readable names to functions within a binary file. We evaluate \system against two established methods: Symlm~\cite{jin2022symlm}, a transformer-based model that leverages context-sensitive, execution-aware code embeddings derived from assembly code, and Asm2vec~\cite{ding2019asm2vec}, which employs random walks on the CFG to sample instruction sequences, and then utilizes the PV-DM~\cite{le2014distributed} model for joint learning of function embedding. It is important to note that function name prediction involves assigning names from a vocabulary of human-readable words; therefore, we report the micro-F1 score for this task.

As shown in Table~\ref{tab: RQ1_funcname}, \system achieves the highest average F1-score of 83.5\%, outperforming both Asm2vec (60.4\%) and Symlm (77.8\%). While Symlm incorporates contextual information as multi-modal inputs, including execution traces and call graph embeddings, in our setting with a limited fine-tuning dataset and demanding train-test split, it does not achieve the same level of performance as \system, which benefits from knowledge transfer from earlier tasks. Notably, \systemzero performs competitively, achieving an F1-score of 69.8\%, surpassing Asm2vec~\cite{ding2019asm2vec}. This further demonstrates the value of pre-trained knowledge embedded in \system.

\subsubsection{Function Similarity Detection}
\begin{figure*}[t]
    \centering
    \includegraphics[width=0.91\textwidth, height=\textheight, keepaspectratio]{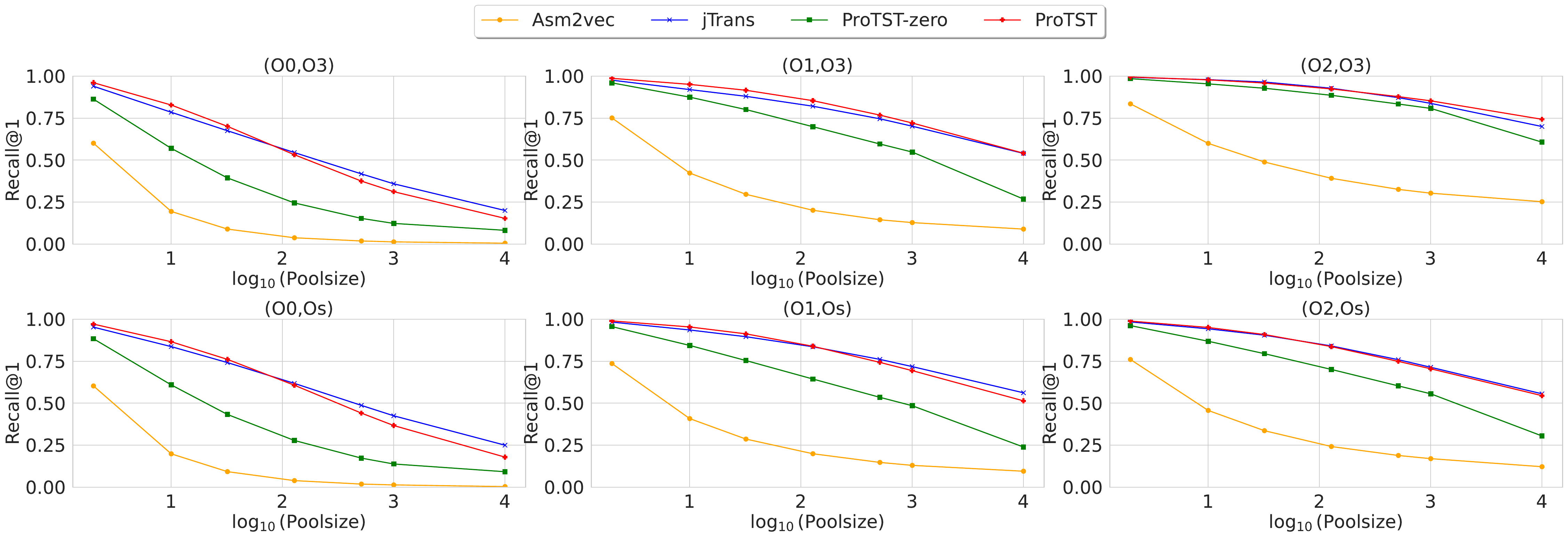}
    \caption{The performance (Recall@1) of different models for binary code similarity detection with respect to pool size.} 
    \label{fig: RQ1_funcsim_recall}
\end{figure*}

The function similarity detection task involves querying a pool of function embeddings to identify the most similar function to a given binary function. We evaluate \system against two established methods, Asm2vec~\cite{ding2019asm2vec} and jTrans~\cite{wang2022jtrans}, a transformer-based model that embeds control flow graphs with assembly code as input.
We use MRR and Recall@1 metrics for different function pool sizes (32 and 10K), as shown in Table~\ref{tab:funcsim}. In Figure~\ref{fig: RQ1_funcsim_recall}, we also present Recall@1 results for various optimization combinations across a broader range of pool sizes (\ie 2, 10, \ldots, 10K). 

Our analysis reveals that the function similarity detection task benefits from a richer vocabulary, granting jTrans an inherent advantage due to its use of assembly code (over 10K tokens~\cite{zhu2023ktrans}) compared to \system's raw byte representation (256 tokens).
This advantage is particularly evident for larger pool sizes. However, even with this limitation, \system achieves competitive results, particularly for smaller function pools (32 or fewer). We observe that \system shows a higher MRR and Recall@1 for small pool sizes in most optimization settings. For larger pool sizes, the performance gap between \system and jTrans narrows to a 2\% difference in average MRR. These results show the effectiveness of the teacher-student learning approach, where transferred knowledge (boundary and function-level knowledge) improves the function similarity detection accuracy. Furthermore, despite Asm2vec directly incorporating function CFG information into its model, \systemzero consistently surpasses in both MRR and Recall@1 across all pool sizes, which shows the benefit of pre-training.

\subsection{RQ2: Generalization of \system} 
In Section~\ref{sec:RQ1}, we demonstrate the effectiveness of \system in various binary analysis tasks and compare it with the state-of-the-art methods.
We now assess its generalization capabilities in two key aspects: ($1$) \system's performance on binaries compiled with different optimization levels (e.g., O0, O1, O2), and ($2$) its effectiveness under diverse code obfuscation techniques.
We benchmark \system against XDA~\cite{pei2020xda} to highlight the distinct advantages of the teacher-student learning paradigm. XDA, a traditional two-stage training approach, serves as a suitable baseline as it operates directly on raw byte sequences. 

\begin{figure*}[t]
    \centering
    \includegraphics[width=\textwidth, keepaspectratio]{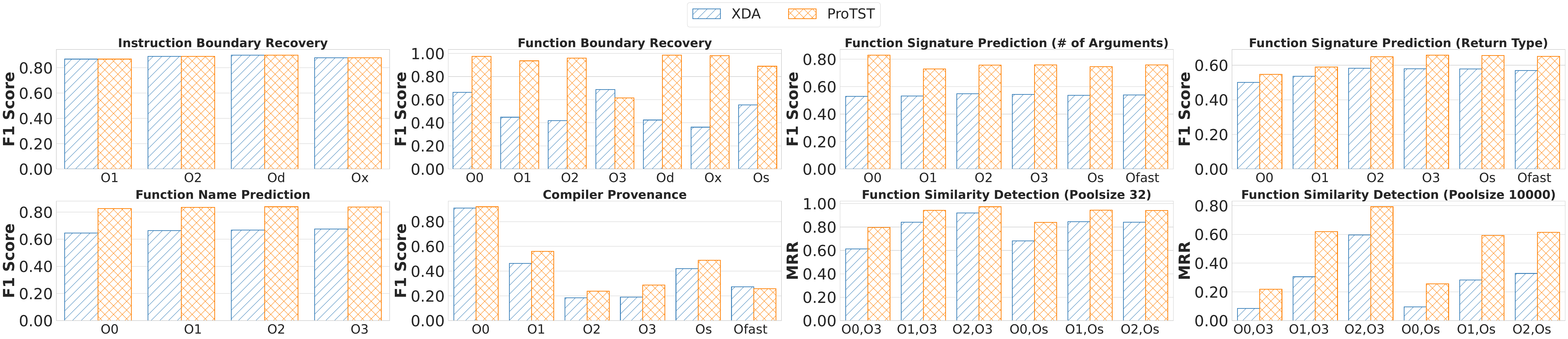}
    \caption{Comparative performance of \system and XDA on various binary analysis tasks across different optimization levels.}
    \label{fig: RQ2_opt}
\end{figure*}

\shortsectionBf{Optimization Levels.} 
To assess \system's ability to address variations in compiler optimization, we categorize binaries based on their optimization flags, ranging from minimal optimization (O0) to aggressive levels (e.g., Ofast). We note that specific optimization levels may vary depending on the compiler and dataset. Due to the absence of optimization information in the BIG2015 dataset, malware classification was excluded from this evaluation. We then evaluated the performance of both \system and XDA in these categorized datasets, with each optimization level (or optimization pair for function similarity detection) undergoing 100 fine-tuning epochs.

The results are presented in Figure~\ref{fig: RQ2_opt}. For instruction boundary recovery, as expected, the initial task in our teacher-student paradigm, both \system and XDA exhibit comparable performance across optimization levels. However, for subsequent tasks, \system shows significantly improved effectiveness compared to XDA, achieving an average performance improvement of 18.4\% at various optimization levels. This substantial improvement is mainly due to the knowledge transfer with teacher-student learning process, which improves the model's ability to generalize to unseen optimization settings.

\begin{figure*}[t]
    \centering
    \includegraphics[width=\textwidth, keepaspectratio]{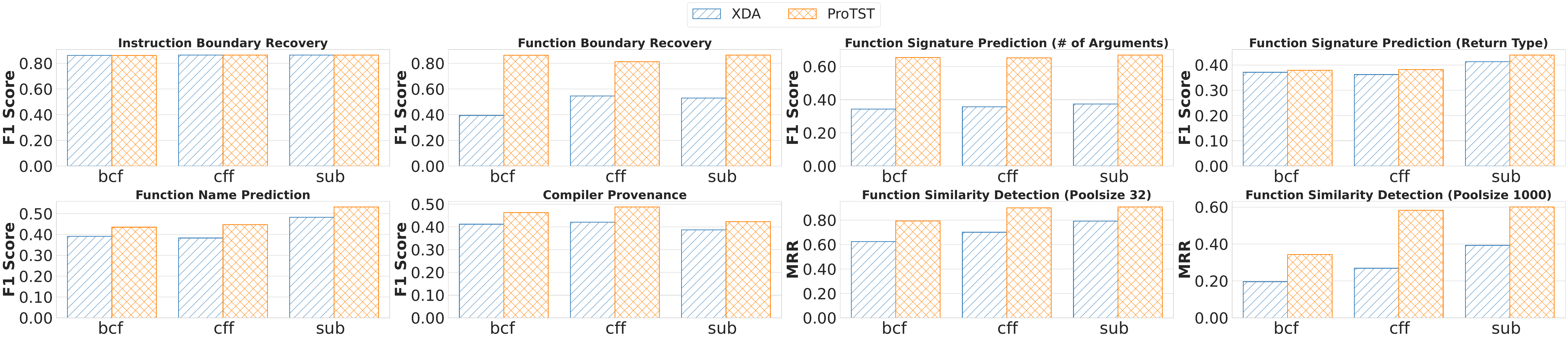}
    \caption{Comparative performance of \system and XDA on various binary analysis tasks under different obfuscation methods.}
    \label{fig: RQ2_obf}
\end{figure*}

\shortsectionBf{Code Obfuscation.} To assess the effectiveness of \system against code manipulation techniques, we evaluate its performance on obfuscated binaries. We used the llvm-obfuscator tool~\cite{junod2015obfuscator} to obfuscate a set of 51 popular open-source software projects (including binutils, curl, and gzip) with three distinct methods: bogus control flow (bcf), instruction substitution (sub), and control flow flattening (cff).

The obfuscation process required several modifications to our initial evaluation setup. First, since only a single obfuscator compiler (llvm-obfuscator) was used, the compiler prediction aspect of the compiler provenance task was omitted, focusing solely on predicting the optimization level. Second, applying obfuscation methods resulted in insufficient complete binary pairs (covering all optimization levels from O0 to Os) for the function similarity detection task with a pool size of 10K. Therefore, we report results using a smaller pool size of 1K to ensure sufficient data for robust evaluation. Third, due to the absence of corresponding source code in malware datasets, which is required by llvm-obfuscator, malware classification was excluded from this evaluation. We leveraged the binary knowledge learned from the pre-trained models in \system, without any pre-training on obfuscated binaries, and performed fine-tuning on the obfuscated data following the strategy outlined in Section~\ref{sec: training config}.

Figure~\ref{fig: RQ2_obf} summarizes the results. Our findings align with those from the optimization-level experiments. We observe that \system outperforms XDA, achieving an average improvement of 16.6\% across all obfuscation methods and tasks. This is a notable improvement considering that \system was not pre-trained on any obfuscated binaries. We postulate that explicitly training the model on obfuscated code at the pre-training stage could further improve its effectiveness against such code manipulation techniques.

\begin{figure}[t!]
    \centering
    \includegraphics[width=1\columnwidth, keepaspectratio]{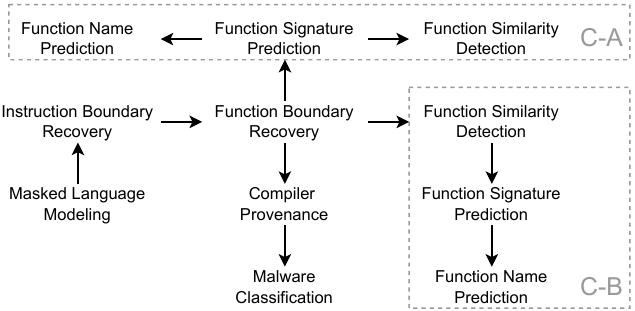}
    \caption{Different knowledge transfer configs (\texttt{C-A} and \texttt{C-B}).} 
    \label{fig: RQ3_two_configurations}
\end{figure}


\subsection{RQ3: Ablation Study on Knowledge Transfer}
\label{sec: RQ2}
We investigate the impact of the staged design in \system's Binary Knowledge Accumulation (BKA) module, focusing on how each preceding task influences the performance of downstream tasks. To investigate this, we conduct ablation experiments to evaluate two configurations of the BKA stage with altered task orders (\texttt{C-A} and \texttt{C-B}), as depicted in Figure~\ref{fig: RQ3_two_configurations}.

Table~\ref{tab: RQ3_BKA_Ablation} summarizes the ablation study results. Due to the similarity between \texttt{C-A} and \texttt{C-B}, only the differing settings and their corresponding results are reported for \texttt{C-B}. We exclude function name prediction and malware classification from evaluation as teacher modules since they are leaf tasks in both configurations. Similarly, function similarity detection, a leaf task in \texttt{C-A}, does not serve as a teacher in that configuration.

In summary, detailed below, our results demonstrate a consistent trend: Each task's performance improves by incorporating prior knowledge from the preceding teacher tasks. As we progress from a lower ID (less prior knowledge) to an adjacent higher ID (more prior knowledge), we observe a minimum performance gain of 2\%. Notably, comparing the best results (bolded in the table) for each task with the traditional two-stage training baseline (ID-1) reveals a consistent improvement of an average 14.8\% with at least 5\% across all downstream tasks.

\begin{table*}[t]
\centering
\def\arraystretch{1.1}
\caption{Analyzing task effectiveness in our design choices. Each row corresponds to a model pre-trained with a specific combination of task modules listed in the ``Teacher Modules'' column. Row $ID-0$ serves as the baseline without pre-training and row $ID-1$ represents the traditional two-stage framework with only MLM pre-training. A dash (-) indicates no result for a specific student task under a particular combination of teacher modules.}
\begin{threeparttable}
\resizebox{\textwidth}{!}{%
\begin{tabular}{c|c|cccccc|ccccccccc}
\hline
 &  & \multicolumn{6}{c|}{\textbf{Teacher Modules}} & \multicolumn{9}{c}{\textbf{Student Tasks}} \\
\multirow{-2}{*}{ID} & \multirow{-2}{*}{Config.} & MLM & Inst Boun & Func Boun & Func Sim & Func Sig & Comp Prov & Inst Boun & Func Boun & Func Sig Count & Func Sig Type & Func Name & Func Sim 32 & Func Sim 10000 & Comp Prov & Mal CLS \\ \hline
\underline{0} &  &  &  &  &  &  &  & 0.871 & 0.478 & 0.529 & 0.495 & 0.641 & 0.768 & 0.257 & 0.368 & 0.714 \\
\underline{1} &  & \ding{51} &  &  &  &  &  & \textbf{0.884} & 0.492 & 0.538 & 0.556 & 0.693 & 0.791 & 0.282 & 0.406 & 0.73 \\
2 &  & \ding{51} & \ding{51} &  &  &  &  & - & \textbf{0.915} & 0.62 & 0.587 & 0.788 & 0.816 & 0.29 & 0.449 & 0.743 \\
3 &  & \ding{51} & \ding{51} & \ding{51} &  &  &  & - & - & 0.626 & 0.613 & 0.805 & 0.827 & 0.332 & \textbf{0.46} & 0.787 \\
4 &  & \ding{51} & \ding{51} & \ding{51} &  & \ding{51} &  & - & - & - & - & 0.814 & 0.865 & 0.393 & - & - \\
5 & \multirow{-6}{*}{A} & \ding{51} & \ding{51} & \ding{51} &  &  & \ding{51} & - & - & - & - & - & - & - & - & \textbf{0.801} \\ \hline
6 &  & \ding{51} & \ding{51} & \ding{51} & \ding{51} &  &  & - & - & \textbf{0.763} & \textbf{0.626} & 0.81 & - & - & - & - \\
7 & \multirow{-2}{*}{B} & \ding{51} & \ding{51} & \ding{51} & \ding{51} & \ding{51} &  & - & - & - & - & \textbf{0.835} & \textbf{0.906} & \textbf{0.515} & - & - \\ \hline
\end{tabular}%
}
    
\end{threeparttable}
\label{tab: RQ3_BKA_Ablation}
\end{table*}


\shortsectionBf{Impact of Task Order.}
We observe that \texttt{C-B} surpasses \texttt{C-A} in function signature prediction, demonstrating notably higher performance for both the prediction of the number of arguments (76.3\% in \texttt{ID-6} compared to 62.6\% in \texttt{ID-3}) and return type (62.6\% in \texttt{ID-6} compared to 61.3\% in \texttt{ID-3}). Similarly, function name prediction also benefits from the task order in \texttt{C-B}, reaching 83.5\% accuracy (\texttt{ID-7}) compared to 81.4\% in \texttt{C-A} (\texttt{ID-4}). This improvement is attributable to the inclusion of function similarity detection as an additional teacher task in \texttt{C-B}, which provides the model with valuable knowledge beneficial for the downstream tasks.

However, \texttt{C-B} presents a trade-off. Although it excels in function signature and name prediction, its performance in function similarity detection is marginally diminished compared to \texttt{C-A}. In \texttt{C-A}, function similarity detection achieves 86.5\% accuracy for a function pool size of 32 (\texttt{ID-4}), and 39.3\% for a pool size of 10K (\texttt{ID-4}). In contrast, when placed earlier in the BKA hierarchy in \texttt{C-B} (\texttt{ID-3}), it receives less knowledge accumulated from previous tasks, resulting in slightly lower performance (82.7\% for a pool size of 32 and 33.2\% for a pool size of 10K). This finding underscores the significance of task order, as tasks positioned later in the hierarchy can leverage accumulated knowledge from teacher tasks and lead to improved performance.

\shortsectionBf{Impact of Cyclical Transfer.}
We investigate cyclical knowledge transfer within the BKA module, where tasks iteratively refine each other's representations. Specifically, we show that one task can enhance the embeddings of another, subsequently improving the original task through this refined knowledge. We illustrate this using function similarity detection (\texttt{ID-7} in \texttt{C-B}), where function signature prediction (\texttt{ID-7}) benefits from having function similarity detection as an additional teacher task compared to \texttt{C-A} (\texttt{ID-4}). The knowledge gained from function similarity detection during pre-training enhances function signature prediction accuracy, which is then transferred back to function similarity detection during fine-tuning. We ensure no data leakage by keeping the fine-tuning and pre-training data separate (Section~\ref{sec: training config}).

Our results demonstrate significant improvements in function similarity detection when it serves as a teacher task for function signature prediction. \texttt{ID-7} outperforms \texttt{ID-4} (pool size 32: 90.6\% vs. 86.5\%; pool size 10K: 51.5\% vs. 39.3\%). This cyclical transfer strategy, applicable to any task configuration, opens avenues for diverse teacher-student combinations, potentially enhancing the performance of all involved tasks.

\begin{figure*}[t!]
    \centering
    \includegraphics[width=\textwidth, keepaspectratio]{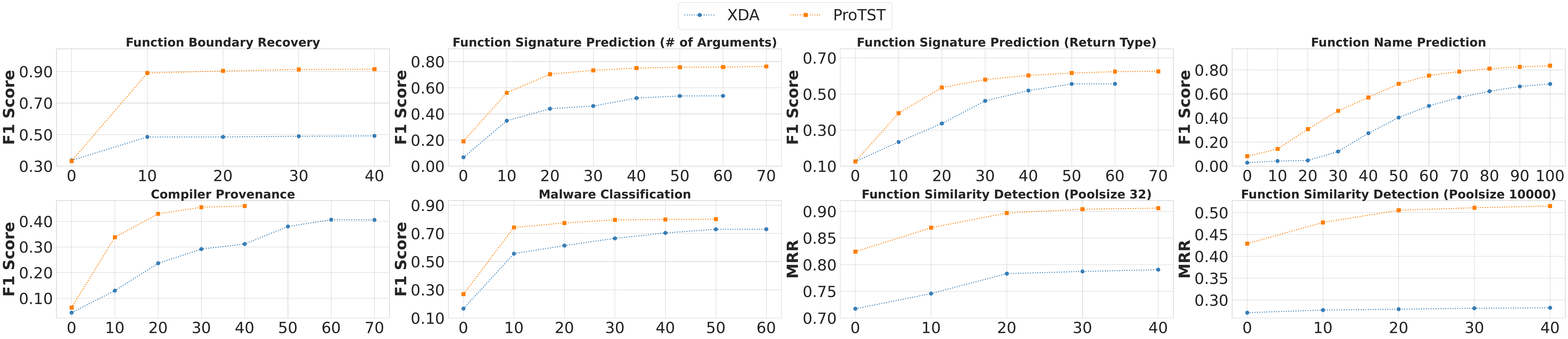}
    \caption{Convergence analysis of validation scores between \system and XDA.} 
    \label{fig: RQ4_epochs}
\end{figure*}

\begin{figure}[ht!]
    \centering
    \includegraphics[width=\columnwidth, keepaspectratio]{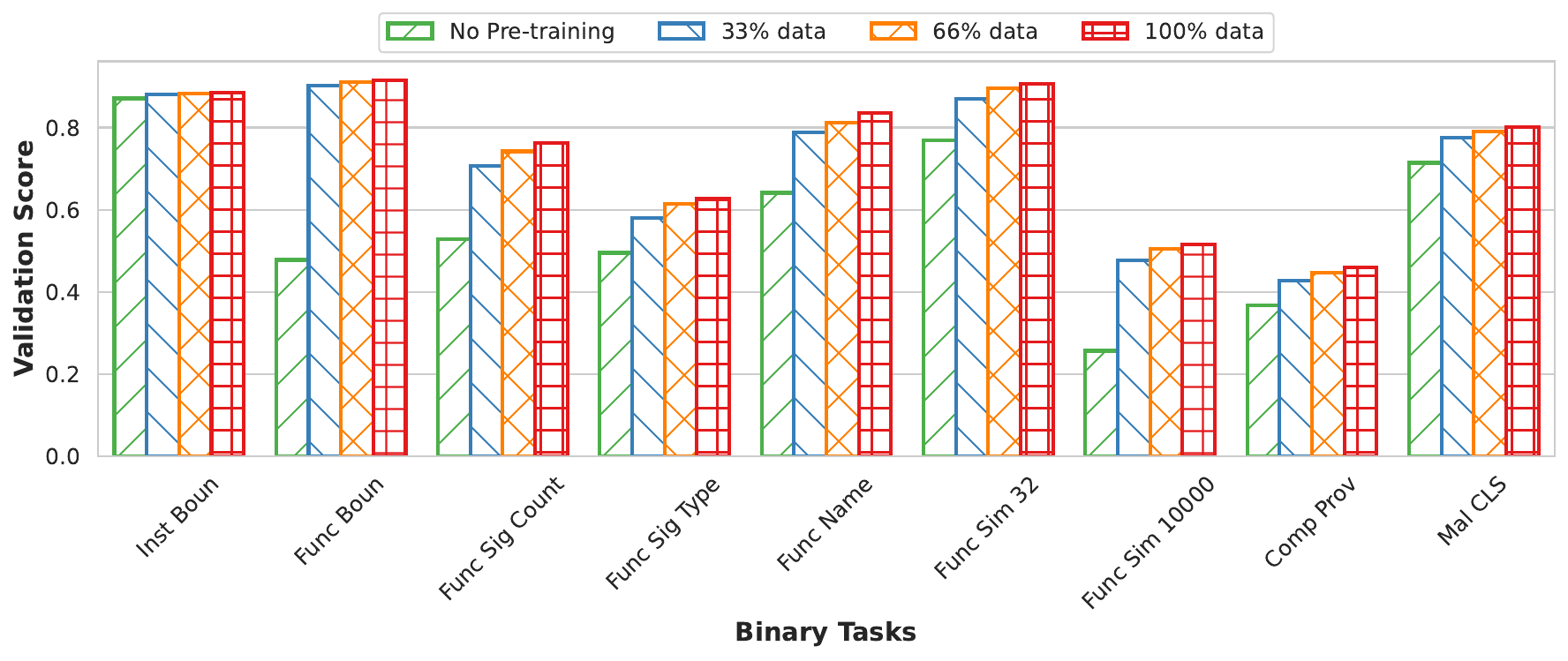}
    \caption{The results of data scaling experiments.} 
    \label{fig: RQ4_data_scaling}
\end{figure}

\subsection{RQ4: Pre-training Effectiveness}
To evaluate \system's efficiency in adapting to new tasks with limited data and fine-tuning epochs, we perform two experiments. First, we analyze the number of fine-tuning epochs required for \system to achieve the desired performance on validation data. This enables us to evaluate how effectively the model can adapt to new tasks with limited fine-tuning. Second, we examine the impact of limiting the number of training samples used in the teacher's tasks during pre-training. This simulates resource constraints and assesses the generalizability of \system to such limitations. We benchmark \system against XDA~\cite{pei2020xda}, a model that lacks progressive knowledge transfer.



\shortsectionBf{Fine-tuning Epochs.} 
We assess whether \system requires fewer fine-tuning epochs than XDA to exceed a chosen performance threshold on the validation data. A lower number of epochs indicates faster convergence and suggests that the pre-trained model is equipped with richer knowledge, which allows for a quicker adaptation to new tasks. We note that the instruction boundary recovery task, being the first task, is expected to exhibit the same convergence speed. Therefore, we exclude this task from the fine-tuning epoch comparison.

Figure~\ref{fig: RQ4_epochs} presents the convergence analysis of the validation scores between \system and XDA.
\system requires fewer fine-tuning epochs to exceed a given threshold than XDA in all downstream tasks. For example, in the compiler provenance task, XDA requires $30$ epochs to reach an F-$1$ score of $0.3$, while \system achieves this in approximately $10$ epochs. Across most binary tasks, it shows an average convergence of two to three times faster than XDA to reach a specific performance threshold. This is because the BKA module accumulates different types of binary knowledge progressively, and the acquired knowledge base facilitates faster convergence during fine-tuning on downstream tasks.

\shortsectionBf{Data Scaling.} To investigate the impact of pre-training data size on the performance of \system, we evaluate four scenarios: The model is fine-tuned directly without any pre-training stage (\texttt{S1}), pre-trained using only $33\%$ of the available data (\texttt{S2}),  pre-trained using only $66\%$ of the available data (\texttt{S3}), and pre-trained using the entire dataset (\texttt{S4}). 


As shown in Figure~\ref{fig: RQ4_data_scaling}, the model performance suffers significantly (with an average drop of $17.6\%$) when no pre-training is applied (\texttt{S1}). This observation highlights the importance of pre-training in equipping the model with foundational knowledge for effective downstream performance. 
Interestingly, we observe that the fine-tuned models achieve a similar validation score. Even when pre-trained with only $33\%$ (\texttt{S2}) or $66\%$ (\texttt{S3}) of the data, the decrease in the validation score compared to the entire dataset pre-training (\texttt{S4}) is minimal, staying within the range $4\%$ for most binary tasks. This suggests that we can potentially achieve similar performance levels while using a reduced amount of pre-training data; this demonstrates the efficiency of \system's pre-training process and its ability to leverage a smaller dataset.

\section{Discussion and Limitations}

Due to computational constraints, we opted for raw byte input, as assembly code possesses a significantly larger vocabulary~\cite{zhu2023ktrans}, which would lead to a significantly longer convergence time in our multi-stage pre-training framework; however, the core knowledge transfer technique of \system remains applicable to assembly code by substituting our raw-byte backbone model with an assembly-based one from previous studies~\cite{jin2022symlm, pei2020trex, wang2022jtrans, zhu2023ktrans, li2021palmtree}. This adaptability enables the incorporation of additional binary analysis tasks, including those based on assembly-level input, into the \system's learning process, potentially enhancing its overall capabilities. Future work will explore tasks such as vulnerability search, indirect call recognition, and memory dependency analysis within \system. 

The hierarchical task order in \system is currently manually determined based on the logical flow of binary knowledge. However, due to computational constraints, the optimal task ordering for \system has not been fully explored. A promising future direction involves leveraging curriculum learning~\cite{bengio2009curriculum} to extend PROTST to new tasks and optimize the task order in a more systematic manner. Additionally, to accommodate an increasing number of tasks efficiently, lightweight fine-tuning approaches such as LoRA~\cite{hu2021lora} or sparse transformers such as Longformer~\cite{beltagy2020longformer} could be employed to enable faster and more scalable fine-tuning.

Although we primarily focus on the x86 architecture due to space constraints, \system is readily applicable to other instruction sets, such as ARM and MIPS, particularly with datasets like~\cite{kim2022revisiting} that include binaries for these architectures. \system requires only the ground truth of binary tasks, which can be readily obtained from labeled datasets or through straightforward manual extraction. Future work will explore the application of \system to these architectures.

\system leverages parameter-based knowledge transfer. Alternative methods such as feature-based and unified-loss learning offer distinct approaches to knowledge transfer. In feature-based transfer learning, each task generates feature embeddings as supplementary input for subsequent tasks within the hierarchical structure. This facilitates the propagation of knowledge from earlier to later tasks through these learned representations.
Unified-loss learning, on the other hand, treats all task-specific loss functions as a single, unified loss, thereby supervising all tasks simultaneously. This approach allows knowledge transfer to benefit from the backpropagation of gradients from later tasks to earlier ones, potentially leading to further refinements in the learned representations. However, both require the design of a specialized model architecture and a training procedure. We are actively investigating the implementation and evaluation of both feature-based and unified-loss within the \system framework.

\section{Conclusion}
We introduce \system, a Progressive Teacher-Student Binary Analysis framework specifically designed to enhance both the accuracy and efficiency of binary analysis tasks. Unlike traditional two-stage training approaches, \system employs a hierarchical tree structure that facilitates a progressive knowledge transfer from fundamental to more specialized tasks. This hierarchical design ensures a natural and logical flow of information, where foundational tasks establish a robust base for more complex tasks. This progressive approach minimizes the reliance on external tools and avoids the tedious processes associated with reverse engineering and feature extraction, thereby simplifying the incorporation of binary code knowledge. Our extensive evaluations underscore the efficacy of \system in a broad range of binary analysis tasks. The results of intensive testing reveal that our progressive teacher-student framework significantly exceeds existing methods regarding learning efficiency and task performance. We believe that our method opens up new avenues for research and offers a promising starting point for future work.






\section*{Acknowledgment}
We thank our shepherd and the anonymous reviewers for their valuable suggestions. This work was supported by NSF under Grant IIS-2229876. Any opinions, findings, and conclusions in this paper are those of the authors and do not necessarily reflect the views of our sponsors.



%
\bibliographystyle{IEEEtran}

\appendices

\section{Binary Task Head Details}
\label{sec: head details}
\system contains different heads for seven different binary tasks integrated into its model architecture. 

\shortsectionBf{Instruction Boundary Recovery Head.} Instruction boundary recovery (\circled{\small{2}}) aims to identify the starting points of the instructions and whether subsequent bytes belong to the same instruction sequence. Give an input byte sequence $x$ of length $n$, the corresponding ground truth $y^{IB}$ is another sequence of labels of the same length $n$:
    \begin{equation}
        y^{IB} = \{y_1^{IB}, y_2^{IB}, ..., y_n^{IB}\}
    \end{equation}
where each element $y_i^{IB}$ in the sequence belongs to the set $\{SI, MI\}$. Here, ``SI'' denotes the ``Start of Instruction'' and ``MI'' denotes ``Middle of Instruction''. Each byte $x_i$ in the input sequence is assigned a corresponding label $y_i^{IB}$ in the ground-truth sequence. To train the model effectively for this token-level classification task, a cross-entropy loss function is employed. This function measures the difference between the predicted probabilities for each byte's class ($SI$ or $MI$) and the actual ground truth labels. Mathematically, it can be represented as:
    \begin{equation}
        \mathcal{L}_{IB} = - \sum_{i=1}^n \sum_{c=1}^C y_{i,c}^{IB} \log P(y_{i,c}^{IB} \mid x_i)
    \end{equation}
where $C$ is the number of classes (2 in this task, $SI$ and $MI$) and $y_{i,c}^{IB}$ is a binary indicator (0 or 1) that is 1 if the true class of the 
i-th byte is $c$. In simpler terms, this objective compares the true class labels for each byte $x_i$ against the predicted probabilities across all possible classes $C$ ($SI$ and $MI$).

\shortsectionBf{Function Boundary Recovery Head.} Function boundary recovery (\circled{\small{3}}) aims to identify the starting point, the continuation within the function, and the end point of functions within binary files. Given an input byte sequence $x$ of length $n$, the corresponding ground truth label $y^{FB}$ is another sequence of labels of the same length $n$. This can be mathematically expressed as:
    \begin{equation}
        y^{FB} = \{y_1^{FB}, y_2^{FB}, ..., y_n^{FB}\}
    \end{equation}
where each element $y_i^{FB}$ in sequence belongs to the set $\{SF, MF, EF\}$. Here, ``SF'' denotes the ``Start of Function'',``MF'' denotes the ``Middle of Function'', and ``EF'' denotes the ``End of Function''. Each byte $x_i$ in the input sequence is assigned a corresponding label $y_i^{FB}$ in the ground-truth sequence. To train the model effectively for this token-level classification task, a cross-entropy loss function is employed. This function measures the difference between the predicted probabilities for each byte's class ($SF$, $MF$ or $EF$) and the actual ground truth labels. Mathematically, it can be represented as:
    \begin{equation}
        \mathcal{L}_{FB} = - \sum_{i=1}^n \sum_{c=1}^C y_{i,c}^{FB} \log P(y_{i,c}^{FB} \mid x_i)
    \end{equation}
where $C$ is the number of classes (3 in this task, $SF$, $MF$ and $EF$) and $y_{i,c}^{IB}$ is a binary indicator (0 or 1) that is 1 if the true class of the 
i-th byte is $c$. In simpler terms, this objective compares the true class labels for each byte $x_i$ against the predicted probabilities across all possible classes $C$ ($SF$, $MF$ and $EF$).

\shortsectionBf{Function Signature Prediction Head.} Function signature prediction (\circled{\small{4}}) aims to identify the number of arguments a function takes and its return data type. Given a binary function represented by its byte sequence $x$, the corresponding ground truth $y^{FS}$ that defines the function's signature. This can be mathematically expressed as:
    \begin{equation}
        y^{FS} = \{ y_n^{FS}, y_t^{FS} \}
    \end{equation}
where $y_n^{FS}$ represents the number of arguments the function takes. It can take values from the set \{1, 2, 3, 4, 5, ``others''\}. The ``others'' category encompasses functions with more than 5 arguments. $y_t^{FS}$ represents the return data type of the function. It can take values from the set \{int, char, void, double, bool, ``others''\}. The ``others'' category encompasses the less common data types. To train the model effectively for this sequence-level classification task, a cross-entropy loss function is used. This function measures the difference between the predicted probabilities for each aspect of the signature (number of arguments and return type) and the actual ground truth labels. Mathematically, it can be represented as:
    \begin{align}
        \mathcal{L}_{FS} = - \bigg[ & \sum_{c=1}^{C_n} y_{n,c}^{FS} \log P(y_{n,c}^{FS} \mid x) \nonumber \\
        & + \sum_{d=1}^{C_t} y_{t,d}^{FS} \log P(y_{t,d}^{FS} \mid x) \bigg]
    \end{align}
where $C_n$ is the number of classes for the number of arguments (\eg 1, 2, 3, 4, 5, "others"), $C_t$ is the number of classes for the return data type (\eg int, char, void, double, bool, "others"). $y_{n,c}^{FS}$ is a binary indicator (0 or 1) that is 1 if the true class of the number of arguments is $c$. $y_{t,d}^{FS}$ is a binary indicator (0 or 1) that is 1 if the true class of the return data type is $d$. This cross-entropy loss function combines the losses for both predictions into a single loss function by summing the individual cross-entropy losses for each attribute.

\shortsectionBf{Function Similarity Detection Head.} Function similarity detection (\circled{\small{5}}) aims to assess the degree of similarity between two code snippets represented in binary code. Given a pair of functions $x_1$ and $x_2$, the backbone model processes each function separately, generating embeddings $E^{x_1}$ and $E^{x_2}$. These embeddings capture the essential characteristics of the respective functions. The model then determines the similarity between the functions. The ground truth label $y$ takes on a value of either 1 or -1 (1 indicates similar functions and -1 indicates dissimilar functions). To calculate the function embedding, the model first averages the individual embeddings generated by the backbone model for each function $E_i^{x}$. This averaged embedding is then fed into a 2-layer Multi-Layer Perceptron (MLP) network. The MLP network further processes the averaged embedding to produce a final output that reflects the predicted similarity between the functions. The final embedding of function can be mathematically expressed as:
    \begin{equation}
        \mathbf{F}(x) = \text{MLP} \left( \frac{1}{N} \sum_{i=1}^{N} \mathbf{E}_i^{x} \right)
    \end{equation}
We use cosine embedding loss to train this task. Given two function embeddings, the loss can be formulated as:
\[
\mathcal{L}_{CE} = 
\begin{cases} 
1 - \cos(\mathbf{F}(x_1), \mathbf{F}(x_2)) & \text{if } y = 1 \\
\max(0, \cos(\mathbf{F}(x_1), \mathbf{F}(x_2)) - m) & \text{if } y = -1 
\end{cases}
\]
where $\cos(\mathbf{F}(x_1), \mathbf{F}(x_2))$ denotes the cosine similarity between the embeddings $\mathbf{F}(x_1)$ and $\mathbf{F}(x_2)$. $m$ is a margin that helps to distinguish dissimilar pairs, typically $0 \leq m \leq 1$. This cosine embedding loss encourages the model to maximize the similarity for similar functions and minimize it for dissimilar ones.

\shortsectionBf{Function Name Prediction Head.} Function name prediction (\circled{\small{6}}) aims at predicting the names assigned to functions within binary code. Given a binary function represented by its byte sequence $x$, the corresponding ground truth $y^{FN}$ is a sequence of words that represents the function name. This can be mathematically expressed as:
    \begin{equation}
        y^{FN} = \{ w_1^{FN}, w_2^{FN}, ..., w_n^{FN} \mid w_i^{FN} \in V\}, 
    \end{equation}
where $w_i^{FN}$ represents an individual word within the sequence of predicted function names. $V$ represents the complete vocabulary of possible function names. 

This vocabulary consists primarily of English words but may also include: ($1$) Developer-chosen terms such as abbreviations, data types (\eg int, float), ($2$) numbers, and ($3$) misspellings. To address challenges such as morphological variations (word form differences) and frequent occurrences of out-of-vocabulary (OOV) words in function names, we follow the pre-processing strategy~\cite{jin2022symlm}, which involves splitting, segmenting, and lemmatizing (converting words to their base form) function names. Additionally, we adjust thresholds to filter out function names containing tokens that appear either too frequently or infrequently in the training data. This helps ensure a more balanced and fair learning process for the model. The task is framed as a multi-class, multi-label classification problem. We employ the BCElogit loss to solve this task. Mathematically, this loss function can be formulated as:
    \begin{align}
        \mathcal{L}_{FN} = - \sum_{i=1}^n \sum_{j=1}^{|V|} \bigg[ & y_{i,j}^{FN} \log \sigma(P(w_{i,j}^{FN} \mid x)) \nonumber \\
        & + (1 - y_{i,j}^{FN}) \log (1 - \sigma(P(w_{i,j}^{FN} \mid x))) \bigg]
    \end{align}
where $n$ is the length of the function name sequence, $|V|$ is the size of the vocabulary $V$. $y_{i,j}^{FN}$ is a binary indicator (0 or 1) that is 1 if the $i$-th word in the function name is the $j$-th word in the vocabulary $V$. $\sigma$ represents the sigmoid function, which converts logits into probabilities. This loss function measures the discrepancy between the predicted probabilities and the actual labels for each word in the function name.

\shortsectionBf{Compiler Provenance Head.} Compiler provenance (\circled{\small{7}}) aims to identify the compiler and optimization level used to generate a binary file. Given a binary file represented by its byte sequence $x$, the model predicts a pair of labels $y^{CP}$. This pair identifies the compiler and the optimization level used for compilation. This can be  expressed as:
    \begin{equation}
        y^{CP} = \{y_c^{CP}, y_o^{CP} \}
    \end{equation}

where $y_c^{CP}$ represents the compiler used. For instance, it can take values from the set \{clang, gcc\}, indicating either the ``clang'' or ``gcc'' compiler. $y_o^{CP}$ represents the optimization level. It takes values from the set \{$O_0$, $O_1$, $O_2$, $O_3$, $O_s$, $O_{fast}$\}, signifying different optimization levels offered by the compilers. To train the model effectively for this sequence-level classification task, a cross-entropy loss function is used. This function measures the difference between the predicted probabilities for each aspect of the signature (compiler and its optimization) and the actual ground truth labels. Mathematically, it can be represented as:
    \begin{align}
        \mathcal{L}_{CP} = - \bigg[ & \sum_{c=1}^{C_c} y_{c,c}^{CP} \log P(y_{c,c}^{CP} \mid x) \nonumber \\
        & + \sum_{o=1}^{C_o} y_{o,o}^{CP} \log P(y_{o,o}^{CP} \mid x) \bigg]
    \end{align}
where $C_c$ is the number of compiler classes (\eg clang, gcc), $C_o$ is the number of optimization level classes (\eg $O_0$, $O_1$, $O_2$, $O_3$, $O_s$, $O_{fast}$). $y_{c,c}^{CP}$ is a binary indicator (0 or 1) that is 1 if the true class for the compiler is $c$. $y_{o,o}^{CP}$ is a binary indicator (0 or 1) that is 1 if the true class for the optimization level is $o$. This loss function combines the errors from the predicted probabilities of both the compiler and the optimization level into a single objective.

\shortsectionBf{Malware Classification Head.} Malware classification (\circled{\small{8}}) aims to categorize a binary file into a specific type of malware family based on its byte sequence. Given a malware file represented by its byte sequence $x$, the model predicts a label $y^{MC}$. This label corresponds to a specific malware family from a predefined list in the BIG2015 dataset~\cite{ronen2018microsoft}. The list includes malware families like Ramnit, Lollipop, Kelihosver3, and others. To train the model effectively for this sequence-level classification task, a cross-entropy loss function is used, which can be formulated as 
    \begin{equation}
        \mathcal{L}_{MC} = - \sum_{k=1}^{K} y_{k}^{MC} \log P(y_{k}^{MC} \mid x)
    \end{equation}
where $K$ is the number of malware family classes (9 in this task, BIG2015~\cite{ronen2018microsoft} contains 9 malware families). $y_{k}^{MC}$ is a binary indicator (0 or 1) that is 1 if the correct class for the malware family is $k$. This loss function evaluates the discrepancy between the predicted probabilities for each malware family and the actual ground truth labels, guiding the model to improve its classification accuracy.

\section{Dataset Details}
\label{sec: dataset}
We present the details of the datasets used to evaluate \system below.

\shortsectionBf{Binutils.} This dataset~\cite{binutils} is generated by compiling the GNU Binutils package with its default settings. It serves as the training data for the MLM pre-training stage.

\shortsectionBf{SPEC CPU.} This dataset includes binaries from various benchmarks (SPEC CPU 2017~\cite{SPEC_CPU_2017} and 2006~\cite{SPEC_CPU_2006}), compiled with different configurations. SPEC CPU 2017 includes 588 binaries, while SPEC CPU 2006 has 333. They are used for instruction and function boundary recovery.

\shortsectionBf{BAP.} This dataset~\cite{bao2014byteweight} contains 2,200 binaries from open-source programs across various platforms (Windows, Linux) and architectures (x86, x64). It's used for function boundary recovery.

\shortsectionBf{BIG2015.} This dataset~\cite{ronen2018microsoft} consists of 10,868 malware samples categorized into nine families. It's used for malware classification.

\shortsectionBf{Binkit.} This dataset~\cite{kim2022revisiting} features 243,128 binaries with 75,230,573 binary functions derived from 51 distinct software packages, compiled using a diverse array of options across compilers and architectures. It's used for tasks like compiler provenance and function signature prediction.

\shortsectionBf{SymLM.} This dataset~\cite{jin2022symlm} includes 16,027 binaries and 1,431,169 functions derived from 27 open-source projects, compiled across multiple architectures and optimization levels using gcc-7.5. It's used for function name prediction.

\shortsectionBf{Binarycorp-3M.} This dataset~\cite{wang2022jtrans} encompasses approximately 3.6 million functions extracted from 10,265 binary programs compiled using gcc and g++ based on ArchLinux packages. It's used for function similarity detection.

\end{document}